\documentclass[twocolumn,twocolappendix,times]{aastex631}
\newcommand{\ngc}{NGC~1514}
\renewcommand{\micron}{\textmu m}
\newcommand{\microns}{\micron}
\newcommand{\rah}{$^{\mathrm h}$}
\newcommand{\ram}{$^{\mathrm m}$}
\newcommand{\eg}{\emph{e.\,g.}}
\newcommand{\kms}{km~s$^{-1}$}
\renewcommand{\deg}{\hbox{$^\circ$}}

\received{September 27, 2024}
\revised{January 24, 2025}
\accepted{February 28, 2025}
\submitjournal{AJ}

\begin{document}

\title{JWST/MIRI Study of the Enigmatic Mid-Infrared Rings in the Planetary Nebula \ngc}

\correspondingauthor{Michael Ressler}
\email{Michael.E.Ressler@jpl.nasa.gov}

\author[0000-0001-5644-8830]{Michael E.\ Ressler}
\affiliation{Jet Propulsion Laboratory, California Institute of Technology,
4800 Oak Grove Drive, Pasadena, CA 91109, USA}
\altaffiliation[]{\textcopyright{} 2025. All rights reserved.}
\author[0000-0003-0884-9589]{Alba Aller}
\affiliation{Observatorio Astron\'omico Nacional (OAN), Alfonso XII 3, 28014, Madrid, Spain}

\author[0000-0003-3947-5946]{David Jones}
\affiliation{Instituto de Astrof\'isica de Canarias, E-38205 La Laguna, Tenerife, Spain}
\affiliation{Departamento de Astrof\'isica, Universidad de La Laguna, E-38206 La Laguna, Tenerife, Spain}
\affiliation{Nordic Optical Telescope, Rambla Jos\'e Ana Fern\'andez P\'erez 7, 38711, Bre\~na Baja, Spain}

 \author[0000-0003-0778-0321]{Ryan M.\ Lau}
 \affiliation{NSF NOIRLab, 950 N. Cherry Ave., Tucson, AZ 85719, USA}

\author[0000-0003-0939-8724]{Luis F.\ Miranda}
\affiliation{Instituto de Astrof\'{\i}sica de Andaluc\'{\i}a (IAA), CSIC, Glorieta de la Astronom\'{\i}a s/n, E-18008 Granada, Spain}

\author[0000-0001-6124-5974]{Karen Willacy}
\affiliation{Jet Propulsion Laboratory, California Institute of Technology, 4800 Oak Grove Drive, Pasadena, CA 91109, USA}

\begin{abstract}
While \ngc{} is an elliptical, but complex, planetary nebula at optical wavelengths, it was discovered to have a pair of infrared-bright, axisymmetric rings contained within its faint outer shell during the course of the WISE all-sky survey. We have obtained JWST mid-infrared imaging and spectroscopy of the nebula through the use of simultaneous observations with the MIRI Imager and Medium Resolution Spectrometer, selecting the F770W, F1280W, and F2550W filters to match each of the MRS's three grating positions. These observations show that the rings are clearly resolved and relatively distinct structures, with both filamentary and clumpy detail throughout. There is also cloud-like material that has a turbulent appearance in the interior of the rings, particularly at the longest wavelengths, and faint ejecta-like structures just outside the ring boundaries. Despite their brightness, the emission from the rings within the three imager passbands is shown to be dominated by thermal emission from very small grains, not line emission from atomic hydrogen or forbidden atomic lines, shocked molecular hydrogen, or PAHs. The doppler velocities derived from the two brightest emission lines in the rings, however, suggest that the material from which the rings were formed was ejected during an early period of very heavy mass loss from the PN progenitor, then shaped by asymmetrical fast winds from the central binary pair.
\end{abstract}


\section{Introduction} \label{sec:intro}

\ngc{} is a large ($\sim 3$\arcmin), moderately high excitation planetary nebula (PN) that lies along the Taurus-Perseus border; the central binary lies at 4\rah{} 9\ram{} 16\fs978 +30\arcdeg 46\arcmin 33\farcs53 (GAIA DR3). While the nebula is still classified as ``elliptical'' at visible wavelengths (\eg{}\ the HASH PN database, \citealt{parker2016}), it is further sub-classified as having ``multiple shells, point symmetry, and internal structure'' since it comprises two shells: a ``lumpy'', bright, inner shell with a significant amount of irregular structure and an attached, diffuse, faint outer shell \citep[\eg][]{chu1987}. However, a pair of axisymmetric, infrared-bright rings was discovered during the course of the Wide-field Infrared Sky Explorer (WISE) all-sky survey \citep[][hereafter R10]{ressler2010} that lie nearly entirely within the outer shell.

Such distinct ring-like structures in PNe are not common, but there are a few examples. Abell 14 (PN A66 14) also has a distinct ring morphology similar in appearance to the WISE images of \ngc{} \citep{akras2016}. However, Abell 14's rings are clearly seen only in the H$\alpha$ and [\ion{N}{2}] narrow-band optical images; our inspection of the WISE catalog images shows a hint of elongated emission inside the rings at W3 (12~\microns{}), but they are undetected at any other wavelength, unlike the situation for \ngc.

There have been several minor refinements in our understanding of \ngc{}'s rings properties since R10 in the absence of additional new infrared data. First is the availability of the GAIA survey data, especially Data Release~3 \citep{gaiadr3}. At the time of R10, the distance to \ngc{} was highly uncertain, ranging from 185~pc from \emph{Hipparcos} data \citep{hip1997} through 1300~pc based on statistical methods \citep[\eg][]{zhang1995}. To set the physical dimensions of the rings, R10 adopted a distance of 250~pc. With the new high-precision value of $454\pm4$~pc from GAIA DR3, we can rescale the diameter of the rings from $\sim 0.2$~pc to 0.4~pc with a separation between the rings scaled up from 0.1\footnote{R10 has a transcription error: the stated separation was the half distance. The full separation between the rings should have been 0.1~pc rather than the stated 0.05~pc.}~pc to 0.2~pc. 

The second refinement is the radial velocity (RV) measurements of \citet{jones2017} that proved that \ngc{} has a binary central source (as opposed to a chance alignment of the white dwarf with an unrelated A0 star) with a mass ratio (cool star / hot star) of order 2.7 with a period of $9.05\pm0.16$~yr, the longest orbital period yet measured in any Galactic PN. These RV measurements of the central binary further support the notion that bipolar PN---in this case we can consider \ngc{} to be bipolar by using the rings to define a polar axis---are formed through the interactions between the central stars as well as with their winds \citep[see the reviews of][and references therein]{jones2017b,boffin2019}, even for widely-separated, long-period binaries. (The orbital parameters determined by \citeauthor{jones2017} imply a semi-major axis of $6.4 \pm 0.7$~AU.) A central binary has also been proposed for Abell 14 to explain its shape, though this has not yet been confirmed observationally.

The third refinement, though perhaps not as important for understanding the nature of the rings themselves, is the optical spectroscopy work, first of \citet{aller2015} who explored the nature of the central source and established that the pair comprised a $T_{\mathit{eff}} \sim 9850$~K post-Main-Sequence A0 star and a $T_{\mathit{eff}} \sim 80,000$ -- 95,000~K sdO star; then of \citet{aller2021} who built a full 3D spatio-kinematical model of the nebula using the {\sc SHAPE} software \citep{steffen2011} based on their high-resolution, long-slit optical spectra of \ngc{}.

Despite these small refinements, our understanding of the nature of the rings has not advanced significantly since their discovery by R10. Because they are prominent only in the mid-infrared (not visible at all at wavelengths shorter than 2~\microns{}---see Appendix \ref{app:palo}), and an informal search of the WISE database for known, resolved PNe by the authors has turned up no similar objects, the ring morphology of \ngc{} appears to be unique. We therefore chose to investigate \ngc{}'s rings in more detail using the Mid-Infrared Instrument (MIRI) on JWST, both through high-spatial-resolution imaging and through spatially resolved medium-resolution spectroscopy where the rings are most prominent.

\section{Observations and Data Processing} \label{sec:obs}

\ngc{} was observed with MIRI on JWST on 2023 Sep 27/28 UTC (JWST Program ID 1238) using simultaneous imaging and spectroscopy. Because the nebula is larger ($\sim 3\arcmin$) than the MIRI imaging field-of-view (1.9\arcmin$\times$1.4\arcmin), and vastly larger than that of the medium-resolution spectrometer (MRS, $\sim$ 3\arcsec--7\arcsec), we chose nine locations around the nebula along with two dedicated MRS backgrounds and one imaging background to sample important locations around the nebula.

\begin{figure*}[ht]
\includegraphics[width=0.33\textwidth]{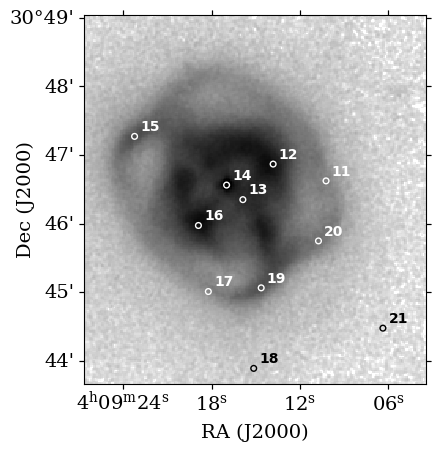}~
\includegraphics[width=0.33\textwidth]{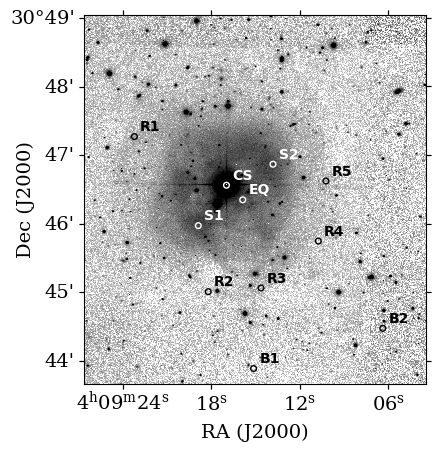}~
\includegraphics[width=0.33\textwidth]{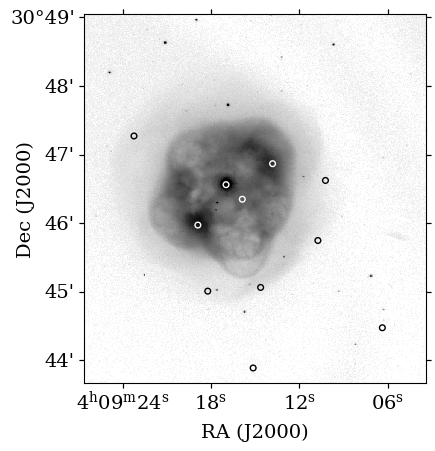}~
\caption{Planning images used for locating the MRS apertures: 12~\microns{} (WISE W3, left), 2.2~\microns{} (Palomar 5-m, WIRC, K$_{\mathrm{short}}$, center), and [\ion{O}{3}] (Calar Alto, details in \citealt{aller2021}). The small circles are 5\arcsec{} in diameter and represent the approximate MRS field-of-view. The observation numbers in the WISE image are those used in the PID 1238 APT file and reflect the observing order that optimizes the ``Traveling Salesman Problem'' to reduce slew overheads. The two-letter codes in the Palomar image are derived from the APT target names (``R'' = rings, ``S'' = shell, etc.) presented in Table~\ref{tab:obs} and the number increases with position angle east of north around the central source. \label{fig:finders}}
\end{figure*}

The pointings were determined by examining WISE images, custom assembled from the low-level WISE L1b data using the ICORE software package \citep{masci2009}, with the PRF-smoothing used in the formal catalog image releases turned off to achieve the highest effective spatial resolution. The MRS positions were chosen to be the central source (Observation 14, see Figure~\ref{fig:finders}, left), a point off the center on the ``equator'' (13), the brightest spot on each of the two inner shell bubbles (12 and 16), and a number of places around the rings (11, 15, 17, 19, and 20). Two additional locations off the nebula (18 and 21) were chosen to serve as the MRS backgrounds. All of these were adjusted so that the simultaneous images would produce a mosaic that covered the majority of the nebula.

Further positional refinements were done using a 2.2~\micron{} image obtained with the Palomar 5-m telescope (observational details in Appendix \ref{app:palo}) and a 5007\AA{} [\ion{O}{3}] image obtained at Calar Alto (observations details in \citealt{aller2021}). At 2.2~\micron{}, the nebulosity is comparatively weak so that foreground/background stars and galaxies are not washed out by the nebulosity. The positions were adjusted so that no known stars or galaxies appeared in the any of the MRS fields-of-view. The [\ion{O}{3}] image allowed us to double-check that the samples in the inner shell were placed at the emission maxima. The naming conventions and observed locations are listed in Table~\ref{tab:obs}.

\begin{table*}[htbp]
\caption{Observations and Coordinates.\label{tab:obs}}
\begin{tabular}{c l c c l}
\hline
Obs \# & Target & RA & Dec & Comment\\\hline
01 & NGC-1514-IMAGE-BKGNG & 04 08 58.182 & +30 49 41.40 & Imager (only) background\\
11 & NGC-1514-RINGS5-W    & 04 09 10.228 & +30 46 37.36 & ``R5'', Northwest ring segment\\
12 & NGC-1514-SHELL2-NW   & 04 09 13.819 & +30 46 52.06 & ``S2'', Bubble in northwestern shell\\
13 & NGC-1514-EQUATOR-SW  & 04 09 15.873 & +30 46 20.95 & ``EQ'', Along ``equator'' of system\\
14 & NGC-1514-CSPN        & 04 09 16.978 & +30 46 33.53 & ``CS'', Central Source \\
15 & NGC-1514-RINGS1-NE   & 04 09 23.226 & +30 47 16.27 & ``R1'', Southeast ring northern edge, opposite \#19\\
16 & NGC-1514-SHELL1-SE   & 04 09 18.890 & +30 45 58.34 & ``S1'', Bubble in southeastern shell\\
17 & NGC-1514-RINGS2-SSE  & 04 09 18.217 & +30 45 00.57 & ``R2'', Southeast ring segment\\
18 & NGC-1514-BKGND1-SSW  & 04 09 15.138 & +30 43 53.34 & MRS background \#1\\
19 & NGC-1514-RINGS3-S    & 04 09 14.631 & +30 45 03.79 & ``R3'', Southeast ring southern edge, opposite \#15\\
20 & NGC-1514-RINGS4-SW   & 04 09 10.743 & +30 45 44.84 & ``R4'', Northwest ring southern edge\\
  21 & NGC-1514-BKGND2-SW   & 04 09 06.372 & +30 44 28.50 & MRS background \#2\\
\end{tabular}

\tablecomments{the Observation numbers, Target names, and coordinates are exactly those used in the JWST Astronomers Proposal Tool (APT) file for Program ID 1238. The typographical error in the ``background'' abbreviation in Observation 01 is also present in the APT file. All observations except 01 used simultaneous MIRI imaging and spectroscopy; 01 used the imager alone. Each observation included all three grating positions to build a complete spectrum, and so for each grating position we took a simultaneous F770W, F1280W, or F2550W image, respectively.}
\end{table*}

\begin{figure*}[ht]
\centering\includegraphics[width=\textwidth]{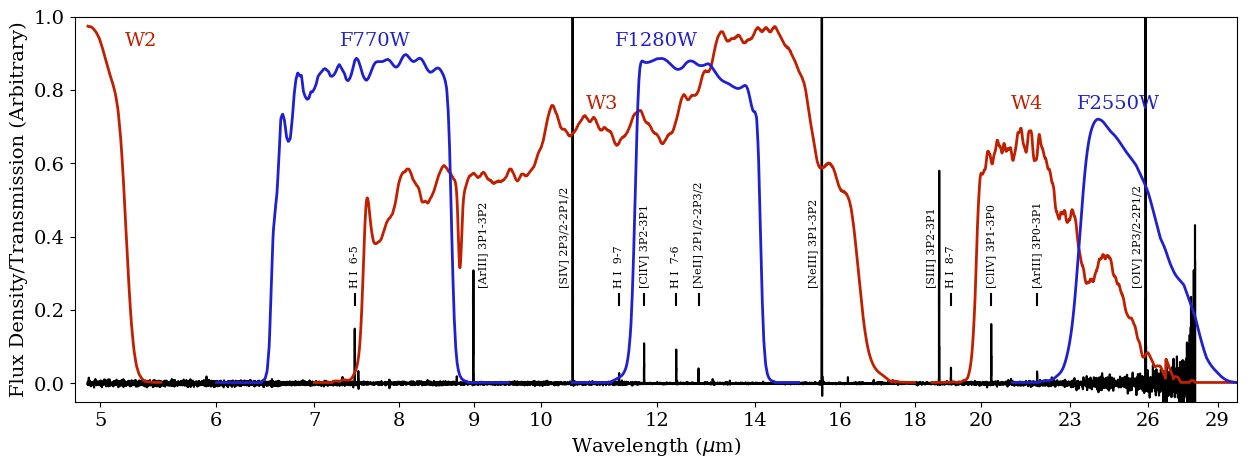}
\caption{The selected MIRI filter passbands (blue) plotted over a continuum-subtracted composite spectrum of the four locations within the inner shell (black). While there are several modest lines seen within the two shorter passbands, they are nearly two orders of magnitude fainter than the [\ion{S}{4}] and [\ion{Ne}{3}] lines we set out to avoid. (On this scale, the [\ion{S}{4}], [\ion{Ne}{3}], and [\ion{O}{4}] lines peak between 15 and 25 units.) Both [\ion{S}{4}] and [\ion{Ne}{3}] are well within the WISE W3 filter (red), potentially contaminating any conclusions about ring temperatures, etc.\ made by R10.\label{fig:nebspecfilt}}
\end{figure*}

Because spectroscopy was the driving consideration in observation planning, the observations were done using an MRS observing template that included simultaneous imaging. Since obtaining a complete MRS spectrum requires three grating positions, we were able to select one imaging filter for each grating position; the F770W and F1280W (7.7 and 12.8~\microns) filters were chosen for their lack of strong emission lines as judged by the Spitzer calibration spectra in R10, and the F2550W (25.5~\micron) filter was selected to get the longest wavelength possible given the rising spectrum seen in R10 (despite the strong [\ion{O}{4}] emission known to be within the passband). The filter passbands for both MIRI and WISE are shown in Figure~\ref{fig:nebspecfilt} plotted over a summed spectrum of the four innermost positions where the emission lines are brightest.

All observations were 238.7 s in duration per position. MRS parameters were always two dither positions with 43 groups in a single integration. The imager data also usually used 43 groups for F770W and F1280W, but for F2550W two integrations of 21 groups per dither were used. Exceptions were for Obs 19 and 20 where the CSPN was in the imager field-of-view; in these two cases, the F770W and F1280W filters also used two integrations of 21 groups.

The data were reprocessed locally using the JWST pipeline v1.16.1 \citep{bushouse2022} and the CRDS context jwst\_1299.pmap. Differences from the standard pipeline processing at Stage 1 involved only cosmic ray handling: we rejected fewer frames after a strike by a ``small'' cosmic ray, but rejected all frames after a cosmic ray ``shower''. At Stage 2, the difference was in background handling; we performed no background subtraction with the standard pipeline. Instead, we created our own background image from Observation 1 by removing some artifacts ``by hand'' after visual inspection, then subtracting these backgrounds from the individual images. We then masked off the Lyot coronagraph portion of each image where they would have overlapped with the main imaging field. In principle the Lyot regions should coadd properly with the main fields, but initial attempts produced linear artifacts in our final coadded images, so we masked them. These images were then coadded using the Stage 3 pipeline by combining all 11 positions using a single association file.

Processing of the MRS spectroscopy proceeded in much the same way. After Stage 2 (with no background subtraction), we masked a number of latent images of extremely bright spectral lines that were left over from previous exposures. These corrected spectra were then coadded and converted to spectral data cubes using the standard Stage 3 pipeline (also with no background subtraction). For the spectra presented here (except for the central source), extractions were done using a median of the entire MRS field-of-view for each wavelength bin, resulting (after some scaling) in surface brightness ``per square arcsecond'' since that is a more appropriate angular scale for the spectroscopy. To create a background, identical extractions were done on the background fields (Observations 18 and 21), then averaged; these 1-D spectra were then subtracted from the rest.

The spectrum of the CSPN was extracted using an aperture that grew as a function of wavelength, set as 2.0 times the effective PSF diameter as described by \citet{law2023}. While this relatively tight diameter may lose some of the flux in the wings of the PSF, it was chosen as a compromise to avoid including too much of the nebular emission in the aperture. The background was derived from a median of eight locations around, but not including, the CSPN from within the same data cube (Observation 14); this very local background includes the nebular emission, but subtracting it from the CSPN should leave (almost) only the flux from the star itself.

\section{Results} \label{sec:results}

\begin{figure*}[ht]
\begin{center}
\includegraphics[width=\textwidth]{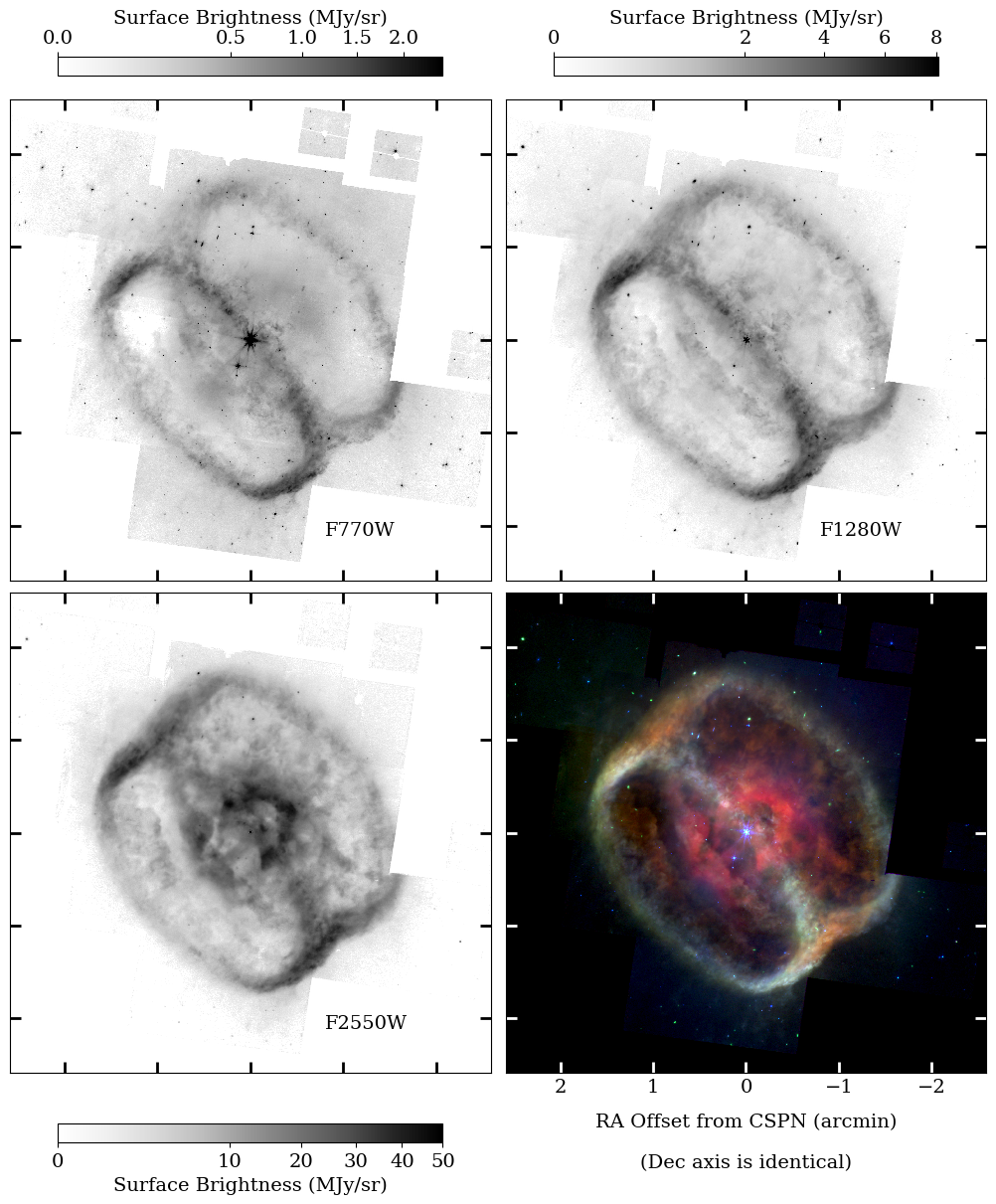}
\end{center}
\caption{Coadded images of \ngc{} at 7.7~\microns{} (upper left), 12.8~\microns{} (upper right), 25.5~\microns{} (lower left), and a three-color composite image of the three. The origin (0\arcmin,0\arcmin) corresponds to the position of the central binary star.
\label{fig:images}}
\end{figure*}

\subsection{Imaging}
\begin{figure}[ht]
\includegraphics[width=\columnwidth]{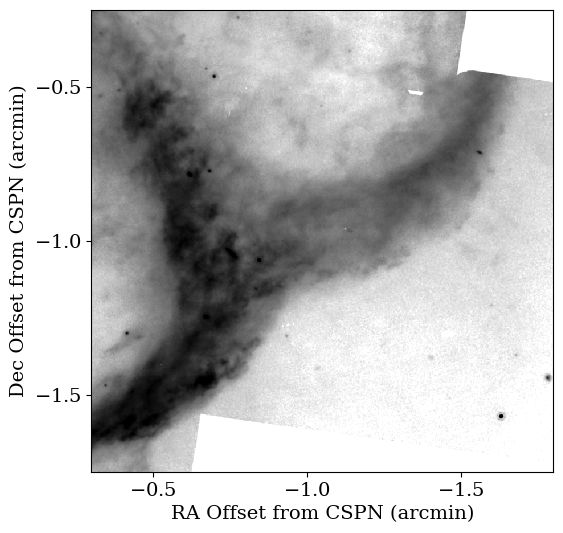}
\caption{Closeup of the rings along their western edges in the F1280W filter. The spatial resolution of MIRI allows us to resolve clumps and filaments in the rings that were smoothed out in the WISE discovery images. Also faintly visible is some material outside the rings (at $\sim$ $-$1.6\arcmin,$-$1.0\arcmin); there is a corresponding material to the northeast on the other side of the rings.\label{fig:closeup}}
\end{figure}

The resulting coadded and registered images are presented in Figure~\ref{fig:images}. Beyond the expected obvious improvement in spatial resolution and sensitivity over WISE, it is striking how bright the rings are with respect to the inner shell at 7.7 and 12.8~\microns, especially the southeast ring. In the WISE data, the rings were fainter than the inner shell at both 4.6 and 12~\microns, but in the MIRI images we find the opposite. The F770W image retains a faint imprint of the optically visible inner shell, but the rings dominate the emission. In F1280W, the shell is not distinctly visible at all; only the rings and a bit of the associated clumpy nebulosity. Figure~\ref{fig:closeup} highlights some of the clumps and filaments visible in F1280W along the western edge of the rings where the rings overlap. While the energetics that formed the rings are clearly dynamic and perhaps a bit chaotic, the environment is still smooth enough that the rings maintain a cohesive structure.

\begin{figure}[ht]
\includegraphics[width=\columnwidth]{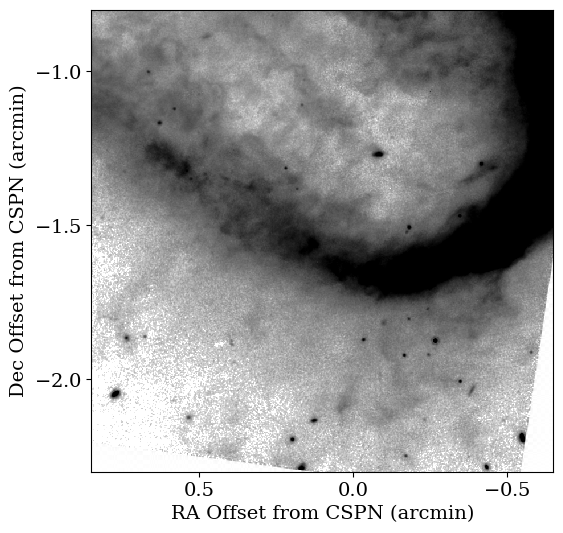}
\caption{Closeup to the south of the southeastern ring showing the ejecta-like features in the F1280W filter. These structures, along with those highlighted in Figure~\ref{fig:closeup}, have not been seen before in either WISE or in optical wavelength imaging.\label{fig:ejecta}}
\end{figure}

The images, though they have very limited field-of-view outside the rings, do indeed show that there is a bit of faint emission that extends outside the ring boundaries at all wavelengths, perhaps ejecta from earlier, less intense outflow activity or from later higher-velocity winds that have passed through the rings. This is especially visible to the south of the southeast ring, approximately (0\arcmin{},$-$2\arcmin) in Figure~\ref{fig:images}, to the southwest of the northwest ring at ($-$1.6\arcmin{},$-$1\arcmin) in Figure~\ref{fig:closeup}, and to the north of the southeast ring at ($+$1.5\arcmin,$+$1.5\arcmin). Figure~\ref{fig:ejecta} shows a closeup of the area to the south of the southeast ring where the extra material is clearest. These extensions are reminiscent of the faint north/south extensions (ansae) seen in NGC 6543 \citep[\eg][]{balick2004} or the northwest/southeast extensions in NGC 7293 \citep[\eg][]{meaburn2005}. A wider field image mosaic is needed to properly characterize these extensions since they extend outside our images (and coincidentally make the background matching in the image registration step more difficult, especially for F770W). However, this emission was not detected in the WISE data nor in any known optical imaging, and so shows that the morphology of the entire nebula is even more complex than understood before.

The F2550W image is perhaps the most striking of all because of its turbulent, flocculent appearance within the ring outlines. It is clear the rings are not completely distinct entities from the rest of the nebula, and that there is not empty space between the rings. There was the barest hint of this in the WISE W4 image, but the new data highlight the chaotic nature of the dust ejecta from the star. 

There is also a bright partial ring (not related to the larger outer rings) contained within the center of the inner shell (the reddest area in the center of the color image in Figure~\ref{fig:images}). We will show later (Section \ref{sec:spectra}) that this ring is composed almost entirely of [\ion{O}{4}] emission, but outside this inner core, the clumpy dust prevails.

\subsection{Spectroscopy\label{sec:spectra}}

\begin{figure*}[ht]
\centering\includegraphics[width=1.0\textwidth]{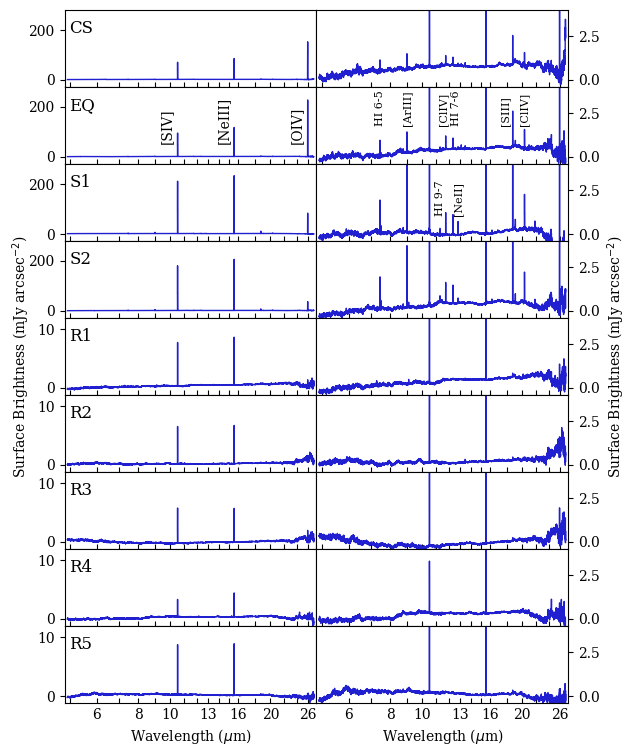}
\caption{Spectra of all sampled locations. The left-hand column shows the spectra at full scale; 0 to 250~mJy~arcsec$^{-2}$ for the four innermost positions, and 0 to 10~mJy~arcsec$^{-2}$ for the ring positions. The right-hand column shows all spectra on a 0 to 3~mJy~arcsec$^{-2}$ scale to highlight the fainter lines. No molecular features, PAHs, or shocked H$_2$ lines are seen at any of the positions. Fitting thermal continua to the spectra was rendered impossible due to the various inflection points caused by the cosmic ray history of both the source observations as well as the two background observations.\label{fig:nebspec}}
\end{figure*}

The spectroscopic data, while helping to clarify what we are seeing (or more often, not seeing) in the images, are surprisingly devoid of features. As in the Spitzer spectra described in R10 (both low- and high-resolution), the MIRI spectra (left column of Figure~\ref{fig:nebspec}) are utterly dominated by the [\ion{S}{4}] emission line at 10.511~\microns, the [\ion{Ne}{3}] line at 15.555~\microns, and the [\ion{O}{4}] line at 25.890~\micron{}. The [\ion{S}{3}] line at 18.713~\microns{} is present as it was for Spitzer, though not nearly as overwhelming as the first three.

\begin{figure*}[ht]
\begin{center}
\includegraphics[width=\textwidth]{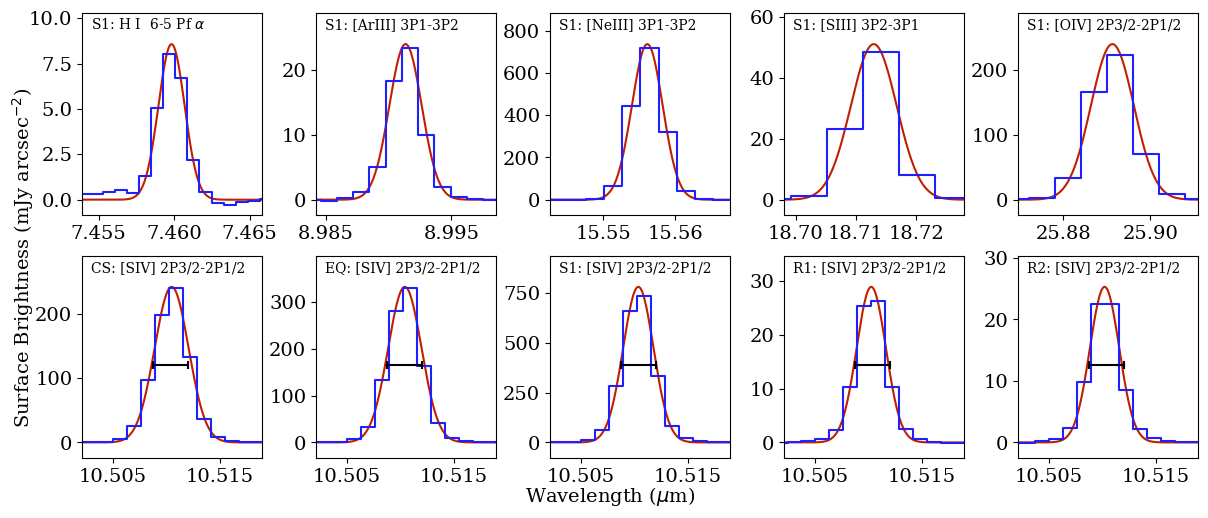}
\end{center}
\caption{Example line fits. The abscissa in all cases ranges from $\lambda - \lambda/1250$ to $\lambda + \lambda/1250$ where $\lambda$ is the tabulated wavelength for the specified line. The top row shows the fits to a number of bright lines spread across the wavelength range of MIRI and partially illustrates the challenge of deriving velocities that will be discussed in Section \ref{sec:disc} due to the changing instrumental line widths. The bottom row shows the fits to the [\ion{S}{4}] line at various positions around the nebula. The black bar at the half power point is derived from the width of the line at R1 and serves as a fiducial so that the velocity broadening, while subtle, can be seen at especially CS and EQ.
\label{fig:line_fits}}
\end{figure*}

Lines that were not detected by Spitzer become visible as one magnifies the flux stretch (right column of Figure~\ref{fig:nebspec}). Various hydrogen recombination lines are visible in emission, as are a number of atomic forbidden lines like [\ion{Ar}{3}], [\ion{Cl}{4}], and [\ion{Ne}{2}]. Even [\ion{Ar}{5}] is faintly visible closest to the CSPN. We measured the surface brightness of each line by first subtracting a local linear continuum, then fitting a gaussian profile to each line (Figure~\ref{fig:line_fits}). Using a more complex profile is not warranted since the FWHM of each profile is almost never more than two spectral slices, so the use of a gaussian profile will provide a consistent estimate of the brightness of each line. The full list of detected lines in each position and estimates of their strengths are given in Table~\ref{tab:lines}.

\begin{table*}[htbp]
  \movetabledown=3.0in
  \begin{rotatetable}
  \caption{Surface brightness and positions for all detected emission lines.}\label{tab:lines}
  \newcommand{\mc}[1]{\multicolumn{2}{c}{#1}}
\begin{tabular}{l D D D D D D D D D D}
\hline
        & \mc{Wavelength} & \multicolumn{18}{c}{Integrated Line Surface Brightness} \\
Species & \mc{(\textmu m)} & \multicolumn{18}{c}{($\times 10^{-20} ~\mathrm{W}\, \mathrm{m}^{-2}\, \mathrm{arcsec}^{-2}$)} \\
        & \mc{} & \mc{CS\tablenotemark{\footnotesize a}} & \mc{EQ} & \mc{S1} & \mc{S2} & \mc{R1} & \mc{R2} & \mc{R3} & \mc{R4} & \mc{R5} \\\hline
\decimals
\ion{H}{1} 9--6 & 5.9082 &   1.31(62) \tablenotemark{\footnotesize b} &   2.86(52)  &   4.74(55)  &   3.29(41)  & \mc{---} & \mc{---} & \mc{---} & \mc{---} & \mc{---} \\
{}[\ion{Na}{3}] 2P1/2--2P3/2 & 7.3177 &   1.79(36)  &   2.01(28)  &   3.07(31)  &   2.27(34)  & \mc{---} & \mc{---} & \mc{---} & \mc{---} & \mc{---} \\
\ion{H}{1} 6--5 (Pf $\alpha$) & 7.4599 &   9.22(46)  &  13.38(29)  &  23.25(36)  &  21.87(32)  &   1.08(28)  &   0.93(29)  & \mc{---} & \mc{---} &   0.52(31)  \\
\ion{H}{1} 8--6 (Hu $\beta$) & 7.5025 &   1.58(28)  &   2.71(23)  &   5.40(23)  &   5.52(27)  & \mc{---} & \mc{---} & \mc{---} & \mc{---} & \mc{---} \\
\ion{H}{1} 10--7 & 8.7601 &   2.05(61)  &   2.80(44)  &   3.02(26)  &   2.54(27)  & \mc{---} & \mc{---} & \mc{---} & \mc{---} & \mc{---} \\
{}[\ion{Ar}{3}] 3P1--3P2 & 8.9914 &  11.86(26)  &  16.24(20)  &  59.64(32)  &  47.04(28)  &   0.56(23)  &   0.66(18)  &   0.44(26)  & \mc{---} &   0.73(17)  \\
{}[\ion{S}{4}] 2P3/2--2P1/2 & 10.5105 & 743.43(3.54)  & 1001.51(5.22)  & 2141.88(8.61)  & 1895.09(7.24)  &  76.48(39)  &  66.81(37)  &  57.63(41)  &  29.66(21)  &  84.48(41)  \\
\ion{H}{1} 9--7 & 11.3087 &   1.32(15)  &   1.70(14)  &   3.13(17)  &   2.72(19)  & \mc{---} & \mc{---} & \mc{---} & \mc{---} & \mc{---} \\
{}[\ion{Cl}{4}] 3P2--3P1 & 11.7619 &   6.15(19)  &   8.59(11)  &  13.49(10)  &  12.35(10)  &   0.46(11)  &   0.49(13)  &   0.36(09)  &   0.30(06)  &   0.46(08)  \\
\ion{H}{1} 7--6 (Hu $\alpha$) & 12.3719 &   4.42(12)  &   6.04(14)  &  10.27(19)  &   9.36(21)  &   0.53(11)  &   0.47(09)  &   0.52(08)  &   0.42(09)  &   0.73(08)  \\
\ion{H}{1} 11--8 & 12.3872 &   0.40(12)  &   0.43(10)  &   0.93(09)  &   1.04(11)  & \mc{---} & \mc{---} & \mc{---} & \mc{---} & \mc{---} \\
{}[\ion{Ne}{2}] 2P1/2--2P3/2 & 12.8135 &   1.36(11)  &   1.44(06)  &   6.47(08)  &   2.71(08)  & \mc{---} & \mc{---} & \mc{---} & \mc{---} & \mc{---} \\
{}[\ion{Ar}{5}] 3P1--3P0 & 13.1022 &   0.99(15)  &   1.71(14)  &   0.34(11)  &   0.10(11)  & \mc{---} & \mc{---} & \mc{---} & \mc{---} & \mc{---} \\
{}[\ion{Ne}{3}] 3P1--3P2 & 15.5551 & 616.73(73)  & 832.51(1.92)  & 1567.80(5.74)  & 1423.83(4.73)  &  53.65(20)  &  43.37(17)  &  37.22(11)  &  27.40(16)  &  58.17(26)  \\
\ion{H}{1} 10--8 & 16.2091 &   0.68(08)  &   0.77(06)  &   1.53(06)  &   1.50(07)  & \mc{---} & \mc{---} & \mc{---} & \mc{---} & \mc{---} \\
{}[\ion{S}{3}] 3P2--3P1 & 18.7130 &  14.62(24)  &  18.94(17)  &  80.98(25)  &  53.79(20)  &   1.35(17)  &   1.44(16)  &   1.33(13)  &   0.11(15)  &   1.39(15)  \\
\ion{H}{1} 8--7 & 19.0619 &   1.49(23)  &   2.42(19)  &   4.42(16)  &   4.20(17)  & \mc{---} & \mc{---} & \mc{---} & \mc{---} & \mc{---} \\
{}[\ion{Cl}{4}] 3P1--3P0 & 20.3107 &   6.10(31)  &   7.99(19)  &  13.07(26)  &  11.76(15)  &   0.36(21)  &   0.38(13)  &   0.28(17)  &   0.11(17)  &   0.59(23)  \\
{}[\ion{Ar}{3}] 3P0--3P1 & 21.8302 &   1.23(23)  &   0.42(25)  &   3.73(31)  &   2.83(28)  & \mc{---} & \mc{---} & \mc{---} & \mc{---} & \mc{---} \\
{}[\ion{O}{4}] 2P3/2--2P1/2 & 25.8903 & 924.54(2.60)  & 1469.24(3.50)  & 492.07(2.44)  & 223.59(1.96)  &   6.02(52)  &   4.80(78)  &   9.87(77)  &   4.85(55)  &   3.23(1.01) \tablenotemark{\footnotesize c} \\
\end{tabular}
  
  \tablenotetext{\qquad\qquad a}{\qquad\qquad The position codes may be found in Figure~\ref{fig:finders} and Table~\ref{tab:obs}.}
  \tablenotetext{\qquad\qquad b}{\qquad\qquad\parbox{7in}{ The uncertainties are given as the two decimal places; \eg{} 1.31(62) means $1.31 \pm 0.62$. For the brightest lines where the uncertainty exceeds 0.99, the full uncertainty is written out, \eg{} 743.42(3.57) means $743.42 \pm 3.57$.}}
  \tablenotetext{\qquad\qquad c}{\qquad\qquad \parbox[t]{8in}{The peak value in the center of this line is ``missing'', so the fit was performed on the wings of this line. This entry may be an underestimate by a factor of $\sim 2$ based on typical line ratios at neighboring positions.}}
\end{rotatetable}
\end{table*}

However, species commonly seen in the mid-infrared in PNe are notably absent in \ngc{}. For example, SMP LMC 058 is a PN that was used for MIRI MRS calibration and it shows clear PAH emission along with SiC \citep{jones2023}. \citet{mata2016} studied a number of PNe with archival Spitzer data, all of which had reasonably prominent H$_2$ 0--0 lines. \ngc{} has none of these. While our spectra do not reliably detect a broad thermal continuum at any of the positions we sampled despite our best attempts to do so, and the baselines are a bit variable due to low signal-to-noise and contamination from residual glow from an unfortunate cosmic ray history on the detectors, emission from any of these molecules still cannot exceed a few hundred \textmu Jy arcsec$^{-2}$ making them negligible contributors to the observed flux.

\subsection{Spectral Mapping}
\begin{figure*}[ht]
  \centering\begin{tabular}{cc}
    \includegraphics[width=0.48\textwidth]{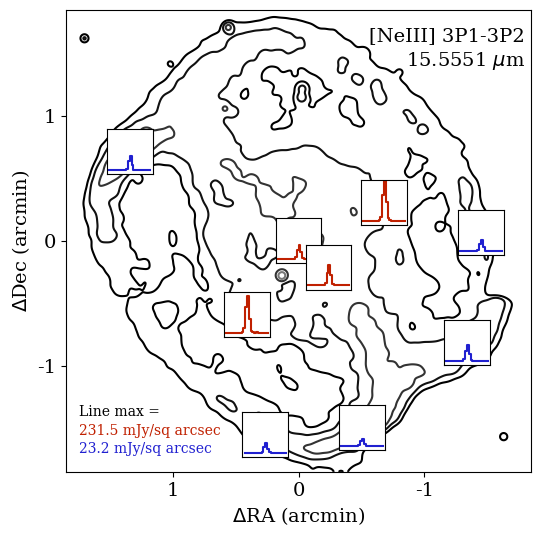} & 
\includegraphics[width=0.48\textwidth]{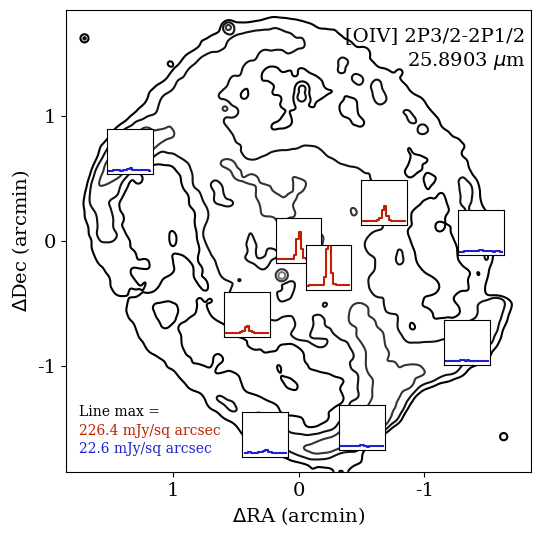} \\
\end{tabular}
\caption{Distribution of [\ion{Ne}{3}] and [\ion{O}{4}] at the observed locations within the nebula. The positions around the rings (blue) are plotted with a y-axis range 1/10th that of the central positions (red). Most detected lines (\eg{} the various H lines when they are detected at all, [\ion{Ar}{3}], [\ion{S}{4}], [\ion{Cl}{4}], [\ion{Ne}{2}], and [\ion{S}{3}]) have a distribution similar to that of [\ion{Ne}{3}]: brightest at the two shell positions, somewhat fainter at the central source, and equator, then an order of magnitude fainter in the rings. Only [\ion{O}{4}] (and perhaps [\ion{Ar}{5}]) is brightest at the center, fainter in the shells, and entirely absent in the rings. \label{fig:specmap}}
\end{figure*}

The intensities of the 15.555~\micron{} [\ion{Ne}{3}] and 25.890~\micron{} [\ion{O}{4}] emission lines are shown for the nine observing locations within the nebula in Figure~\ref{fig:specmap} . (The quoted intensity of the lines at the central binary star do not include emission directly from the binary; only the immediately surrounding nebulous emission.) While these two lines are by far the brightest (along with the [\ion{S}{4}] line at 10.511~\microns), almost all the detected lines (Table~\ref{tab:lines}) follow the pattern seen in the [\ion{Ne}{3}] line at left: brightest at the two positions on the inner shell bubbles (S1 and S2); somewhat fainter but still strong at the CSPN position and the nearby equatorial position (CS and EQ); and an order of magnitude lower at the ring positions (R1--R5). [\ion{O}{4}] is the only significant exception. In this case, the line is brightest at the two central positions (CS and EQ; EQ samples the F2550W inner ring mentioned earlier), fainter in the inner shell, and almost entirely absent out in the rings. [\ion{Ar}{5}] is perhaps a second example since it is detected only weakly at the two central positions and not at all elsewhere. It and [\ion{O}{4}] have ionization potentials of 60 and 55 eV, respectively, the highest of any of the detected lines, and are thus found in the most intense UV fields nearest the stars.

\begin{figure}[ht]
\includegraphics[width=\columnwidth]{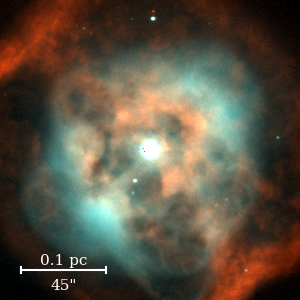}
\caption{Combined image of the inner shell of a visible wavelength [\ion{O}{3}] image with the F2550W image. The relative colors are only qualitative, chosen to highlight the location of the [\ion{O}{4}] ``ring'' close to the CSPN, but [\ion{O}{3}] is mostly the pale blue, and [\ion{O}{4}] in the inner shell is orange. (Recall that the outer rings contain no [\ion{O}{4}]; they have only dust emission.)\label{fig:o3overlap}}
\end{figure}

If we therefore accept that the inner shell emission in the F2550W filter is almost all [\ion{O}{4}] (see Section \ref{sec:ringem}), we can use that to trace the location of [\ion{O}{4}] with respect to the [\ion{O}{3}] seen at visible wavelengths in the inner shell. The [\ion{O}{3}] image shows a number of bubbles in the shell; the [\ion{O}{4}] is seen tracing the interior regions of these bubbles (Figure~\ref{fig:o3overlap}). This suggests that the high energy photons needed to triply ionize oxygen (55 eV) are absorbed at the inner, exposed edges of the bubbles, while those needed to doubly ionize oxygen (35 eV) are able to penetrate deeper into the gas contained in the inner shell.

\subsection{Comparison with WISE Images}
\begin{figure*}[ht]
\centering\includegraphics[width=\textwidth]{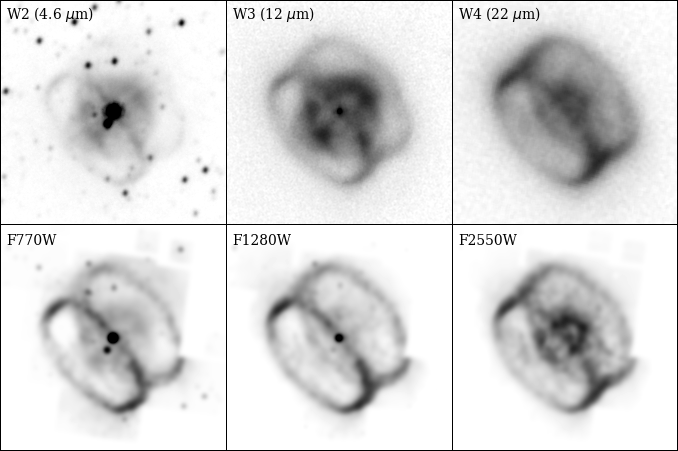}
\caption{Comparison of WISE images (custom-coadded with no PRF smoothing) with blurred, resampled MIRI images meant to mimic the WISE image properties.\label{fig:wisecomp}}
\end{figure*}

Figure~\ref{fig:wisecomp} directly compares the WISE discovery images to blurred and resampled MIRI images, meant to replicate the WISE images as closely as possible. While the F770W image might be thought of lying between W2 and W3 in wavelength space, the implication based on the WISE filters would be that the inner shell should be brighter than the rings, which it clearly is not. The appearance of the inner shell in the F1280W image is nothing like the W3 image even though the centers of the passbands are very close.

Given the spectroscopic evidence above, we can conclude that the difference between the shorter wavelength WISE and MIRI images is the presence or absence of the [\ion{S}{4}] and [\ion{Ne}{3}] lines within the passband. Emission from the optically bright inner shell seems to be dominated by the ionic lines in the mid-infrared, so their absence (by observing program design) explains the faint inner shell at F1280W. The inner shell is slightly more prominent in F770W where there is also no strong ionic emission, but there is some emission from the \ion{H}{1}~6--5 line at 7.456~\microns{}; we discuss this more quantitatively in Section \ref{sec:ringem}.

The blurred F2550W image, on the other hand, is quite a good match to the W4 image, implying they are sampling the same emission sources. While the [\ion{O}{4}] line would seem to lie mostly outside the W4 passband plotted in Figure~\ref{fig:nebspecfilt}, \citet{brown2014} give evidence to suggest that the W4 passband is somewhat redder than the formal spectral response curve, meaning that W4 is more sensitive to the [\ion{O}{4}] line than predicted. If so, it can easily account for the similarity between W4 and F2550W.

Therefore, we conclude that the brightness of the inner shell in the 25.5~\micron{} image is due to the presence of the [\ion{O}{4}] line. Given that the intensity drops so rapidly with distance from the CSPN, we can surmise that if we could remove the [\ion{O}{4}] flux from the F2550W filter, the inner shell would be no brighter with respect to the rings than in the other two MIRI filters.

\subsection{Spectrum of the Central Binary}

\begin{figure}[ht]
\centering\includegraphics[width=\columnwidth]{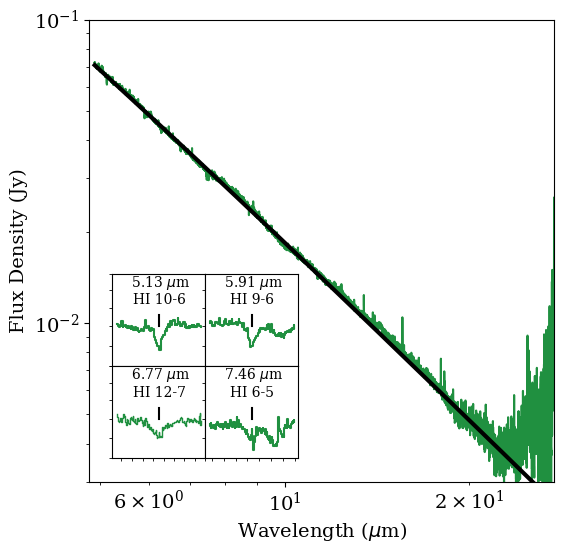}
\caption{Spectrum of the central source. The black line represents a 9730~K pure blackbody fit to the data up to 20~\microns. This fit represents the majority of the spectrum adequately well. The long wavelength tail can be fit by a 50~K blackbody, but this is likely to be incompletely corrected thermal background from the $\sim$~50~K telescope. Insets showing the relative absorption depth of four hydrogen lines are given in the lower left; the continuum level is approximately 1.0, while the lower end of the scale is 0.9, so the depths are approximately 5\%. The dip in the spectrum to the right of \ion{H}{1}~9--6 is from \ion{H}{1}~14--7, and to the right of \ion{H}{1}~6--5 is \ion{H}{1}~8--6 and \ion{H}{1}~11--7. \label{fig:cspnspec}}
\end{figure}

As part of the overall spectral extraction process, we also obtained the spectrum of the CSPN (Figure~\ref{fig:cspnspec}). As in the low-resolution Spitzer spectrum in R10, the emission is dominated by the cooler A0 companion star. The heavy black line in Figure~\ref{fig:cspnspec} shows the best fit blackbody of 9730~K between 5 and 20~\microns, and it provides an adequate fit out to 21 or 22~\microns{} where the spectrum begins to rise. This rise can be fit by a $\sim$~50~K blackbody, but that particular temperature makes it less likely to be astrophysical, but rather that the rise is due to an imperfectly subtracted 50~K telescope background. 

Unlike the Spitzer spectrum obtained at low resolution in R10, we do not see the 100~K dust continuum in our CSPN spectrum. This is due to a combination of a much smaller beam size ($<$~1~arcsec diameter vs a 3.7$\times$57~arcsec slit), much better spatial resolution (0.4 vs 2.5 arcsec), and a background derived from positions only 1.5 arcsec from the CSPN, rather than arcminutes away. For all intents and purposes, the mid-infrared spectrum of the CSPN (which is dominated by the cooler A0 star) simply looks like a hot ($\sim 10000$~K best fit) blackbody. This new result does rule out a circumstellar/binary dust disk like those seen in some PN such as NGC 7293 \citep[][with further examples with and without excesses in \citealt{bilikova2012}]{su2007}; an equivalent disk in \ngc{} would have been easily seen as an excess in this spectrum.

Somewhat reassuringly, however, we do see broadened hydrogen absorption lines in the CSPN spectrum, as one would expect for an A star. The insets in Figure~\ref{fig:cspnspec} show four of the broadened hydrogen lines, though for \ion{H}{1}~6--5 (Pf $\alpha$), we can see the narrow nebular emission line imprinted on the broad photospheric absorption line. At longer wavelengths, the nebular emission dominates over the weakening photospheric absorption, so that just the emission lines remain.

\section{Discussion} \label{sec:disc}

\subsection{Velocity Structure}

The process of fitting the emission lines to derive the surface brightnesses also returned the line widths and line centers. While tabulating the fit results, it became obvious that the lines were widest at the two central positions, less wide but still resolved at the inner shell positions, and narrowest (and presumably unresolved) at the ring positions. Table~\ref{tab:widths} lists the as-measured widths for the stronger lines that have a good SNR. Unfortunately, only the [\ion{S}{4}] and [\ion{Ne}{3}] lines are bright enough to obtain reliable velocity measurements in the rings (see Section \ref{sec:expvel}).

\begin{table*}[htbp]
  \caption{Measured Line Widths (including instrument resolution)\label{tab:widths}}
\newcommand{\mc}[1]{\multicolumn{2}{c}{#1}}
    \hspace*{-1.3in}\begin{tabular}{c DDD DDD DDD}
    Species & \mc{[\ion{O}{4}]} & \mc{[\ion{S}{4}]} & \mc{[\ion{Ne}{3}]} & \mc{[\ion{S}{3}]} & \mc{[\ion{Ar}{3}]} & \mc{\ion{H}{1} 6--5} & \mc{[\ion{Cl}{4}]} & \mc{[\ion{Cl}{4}]} & \mc{\ion{H}{1} 7--6}\\
    Wavelength (\microns)& \mc{25.8903} & \mc{10.5105} & \mc{15.5551} & \mc{18.7130} & \mc{8.9914} & \mc{7.4599} & \mc{11.7619} & \mc{20.3107} & \mc{12.3719}\\
    Slice Width (km s$^{-1}$) & \mc{69.47} & \mc{37.07} & \mc{48.18} & \mc{96.11} & \mc{43.34} & \mc{32.15} & \mc{63.71} & \mc{88.55} & \mc{60.57}\\\hline
    &  \mc{Width} &  \mc{Width} &  \mc{Width} &  \mc{Width} &  \mc{Width} &  \mc{Width} &  \mc{Width} &  \mc{Width} &  \mc{Width} \\
    Pos & \mc{(km s$^{-1}$)} & \mc{(km s$^{-1}$)} & \mc{(km s$^{-1}$)} & \mc{(km s$^{-1}$)} & \mc{(km s$^{-1}$)} & \mc{(km s$^{-1}$)} & \mc{(km s$^{-1}$)} & \mc{(km s$^{-1}$)} & \mc{(km s$^{-1}$)} \\\hline\hline
    \decimals

        CS & 143.5(0.6) & 106.8(0.7) & 107.2(0.2) & 152.9(3.6) & 114.1(3.5) & 92.3(6.6) & 133.1(5.7) & 142.6(11.8) & 115.8(4.4) \\
EQ & 141.8(0.5) & 102.9(0.7) & 103.5(0.1) & 148.9(2.2) & 112.2(2.0) & 94.5(3.0) & 123.3(2.2) & 135.2(4.4) & 119.4(3.8) \\\hline
S1 & 138.0(0.8) & 97.4(0.5) & 99.4(0.2) & 142.5(0.6) & 99.5(0.7) & 87.2(1.8) & 118.0(1.2) & 118.7(3.8) & 106.1(2.5) \\
S2 & 143.6(0.8) & 96.8(0.5) & 98.6(0.3) & 142.7(0.8) & 99.1(0.8) & 86.2(1.7) & 118.0(1.3) & 123.6(2.3) & 105.2(2.9) \\\hline
R1 & \mc{~} & 92.8(0.7) & 95.4(0.3) & \mc{~} & \mc{~} & \mc{~} & \mc{~} & \mc{~} & \mc{~} \\
R2 & \mc{~} & 94.1(0.7) & 95.8(0.3) & \mc{~} & \mc{~} & \mc{~} & \mc{~} & \mc{~} & \mc{~} \\
R3 & \mc{~} & 90.6(0.9) & 93.6(0.4) & \mc{~} & \mc{~} & \mc{~} & \mc{~} & \mc{~} & \mc{~} \\
R4 & \mc{~} & 92.6(0.9) & 95.1(0.7) & \mc{~} & \mc{~} & \mc{~} & \mc{~} & \mc{~} & \mc{~} \\
R5 & \mc{~} & 91.9(0.6) & 96.6(0.2) & \mc{~} & \mc{~} & \mc{~} & \mc{~} & \mc{~} & \mc{~} \\
\end{tabular}
\end{table*}

Table~\ref{tab:widths} lists the lines in order of decreasing brightness at the central positions along with the velocity width of each spectral slice for each of those lines. The FWHM of most of the lines cover only 2 to 3 slices, making it difficult to obtain well-resolved line widths. However, the lines are consistently wider at CS and EQ than S1 and S2, except for [\ion{O}{4}] where the widths are all approximately equal.

If we suppose that the line widths at the ring positions are approximately the instrumental line widths, we can subtract those in a root-sum-square fashion ($v_{true} = \sqrt{ v_{meas}^2 - v_{inst}^2}$) for the [\ion{S}{4}] and [\ion{Ne}{3}] lines to estimate their velocity dispersions. These are given in Table~\ref{tab:corrwidths}; the instrumental widths were calculated from the median of the five ring positions, then subtracted. Dashes are given where the ring line width is less than the median. In round numbers the line widths at CS are about 52~\kms, at EQ 44~\kms, and at the shell positions about 30~\kms. This is broadly consistent with the visible wavelength observations where the doppler velocities of the inner shell are approximately 25~\kms{} \citep[\eg][]{muthu2003,aller2021}. In the center, where we will see emission from both approaching and receding gas, at our spectral resolution we might expect a line width about twice the expansion velocity, about 50~\kms{} in this case. In the inner shell positions, there will still be a mixture of velocities along the line-of-sight, but with one predominant thus narrowing the apparent width. So at least for the optically visible portion of the nebula, our velocity measurements are broadly consistent with the previous measurements.

\begin{table*}[htbp]
  \caption{Corrected Lines Widths for [\ion{S}{4}] and [\ion{Ne}{3}]\label{tab:corrwidths}}
\begin{center}
  \newcommand{\mc}[1]{\multicolumn{2}{c}{#1}}
    \begin{tabular}{c DDDD}
    Species                   & \mc{[\ion{S}{4}]}  & \mc{[\ion{Ne}{3}]} & \mc{[\ion{S}{4}]}  & \mc{[\ion{Ne}{3}]} \\
    Wavelength (\microns)     & \mc{10.5105}       & \mc{15.5551}       & \mc{10.5105}       & \mc{15.5551}       \\
    Slice Width (km s$^{-1}$) & \mc{37.07}         & \mc{48.18}         & \mc{37.07}         & \mc{48.18}         \\\hline
                              & \mc{Raw}           & \mc{Raw}           & \mc{Corrected}     & \mc{Corrected}     \\
                              & \mc{Width}         & \mc{Width}         & \mc{Width}         & \mc{Width}         \\
    Pos                       & \mc{(km s$^{-1}$)} & \mc{(km s$^{-1}$)} & \mc{(km s$^{-1}$)} & \mc{(km s$^{-1}$)} \\\hline\hline
    \decimals

        CS & 106.8(0.7) & 107.2(0.2) & 54.4(2.3) & 49.6(1.4) \\
EQ & 102.9(0.7) & 103.5(0.1) & 46.3(2.7) & 40.9(1.6) \\
S1 & 97.4(0.5) & 99.4(0.2) & 32.5(3.6) & 29.0(2.3) \\
S2 & 96.8(0.5) & 98.6(0.3) & 30.5(3.9) & 26.1(2.7) \\
R1 & 92.8(0.7) & 95.4(0.3) & 13.2(9.1) & 7.9(8.8) \\
R2 & 94.1(0.7) & 95.8(0.3) & 20.3(6.0) & 12.1(5.7) \\
R3 & 90.6(0.9) & 93.6(0.4) & \mc{---} & \mc{---} \\
R4 & 92.6(0.9) & 95.1(0.7) & 11.8(10.8) & \mc{---} \\
R5 & 91.9(0.6) & 96.6(0.2) & \mc{---} & 17.2(3.9) \\
\end{tabular}
\tablecomments{assumed instrumental line widths based on the measurements at the ring positions: [\ion{S}{4}] width $= 92.4 \pm 1.7$ km s$^-1$, [\ion{Ne}{3}] width $= 95.3 \pm 1.0$ km s$^-1$}
\end{center}
\end{table*}

\subsection{Expansion Velocity\label{sec:expvel}}

Since the emission lines in the rings are apparently unresolved, we rely on the small doppler shifts we find from the line profile fitting. However, the velocities derived directly from the fits show significant systematic deviations due to the coarse spectral sampling of the lines ($\sim$ 50--100~\kms{} per spectral channel), small uncertainties in the rest wavelengths of the lines ($\pm 0.0001$~\microns{} yields a $\pm 1$--3~\kms{} change depending on wavelength), and perhaps some residual uncertainties in the spectral calibration of MIRI. To provide a better estimate of the ensemble line velocity at each ring position, we determined an offset correction for each line based on the values derived from the four inner shell positions. The procedure itself is described in Appendix \ref{app:velo}. 

Table~\ref{tab:vel} lists the velocities of the [\ion{S}{4}] and [\ion{Ne}{3}] lines before and after correction. While the correction for [\ion{S}{4}] is relatively small, only 3~\kms{} or so, the systematic offset of [\ion{Ne}{3}] is rather large, resulting in a correction of nearly 20~\kms. After correction, the velocities inferred for each position are reasonably consistent between the two lines, and it appears that there may be some organized motion.

\begin{figure}[ht]
\includegraphics[width=\columnwidth]{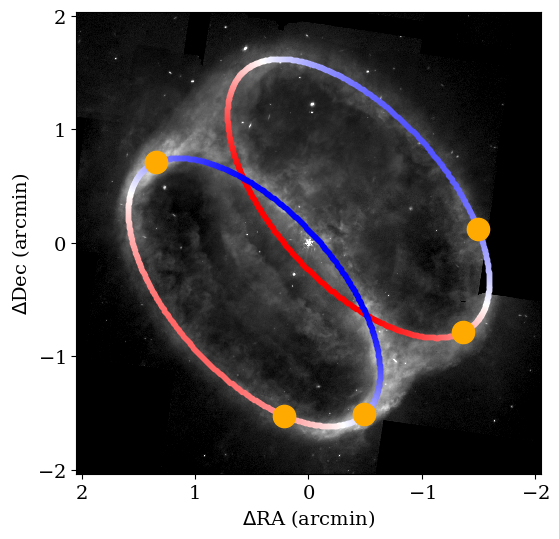}
\caption{Toy model of the rings showing red- and blue-shifted areas if the rings are expanding from the central source. The orange dots show the MRS pointings. Position angles around the rings (not that on the sky) have their origins at the point of each ring closest to us (the bluest part of the model rings, \eg{} near the central source on the SE ring) and increase in a clockwise direction around the ring from our perspective.\label{fig:velmod}}
\end{figure}

Given that the inner shell is already known to be expanding, we build an extremely simple geometrical model of the rings to see if the observed velocities are also consistent with expansion. We model the southeast ring (SER) as an 87\arcsec{} radius circle and the northwest ring (NWR) as a 90\arcsec{} circle. (The diameters were remeasured in this work, but are consistent with R10). The opening angle from the equatorial plane to the SER is approximately 31\deg, while that to the NWR is $-$28\deg. The tilt from edge-on is 30\deg{} (31\deg{} in R10), while the rotation in the plane of the sky is 132\deg{} east of north (131\deg{} in R10). Figure~\ref{fig:velmod} shows the geometry along with the predicted relative red and blue shifts if the rings are expanding radially from the central binary. The positions of the five ring observations are marked with orange dots.

\begin{figure}[ht]
\includegraphics[width=\columnwidth]{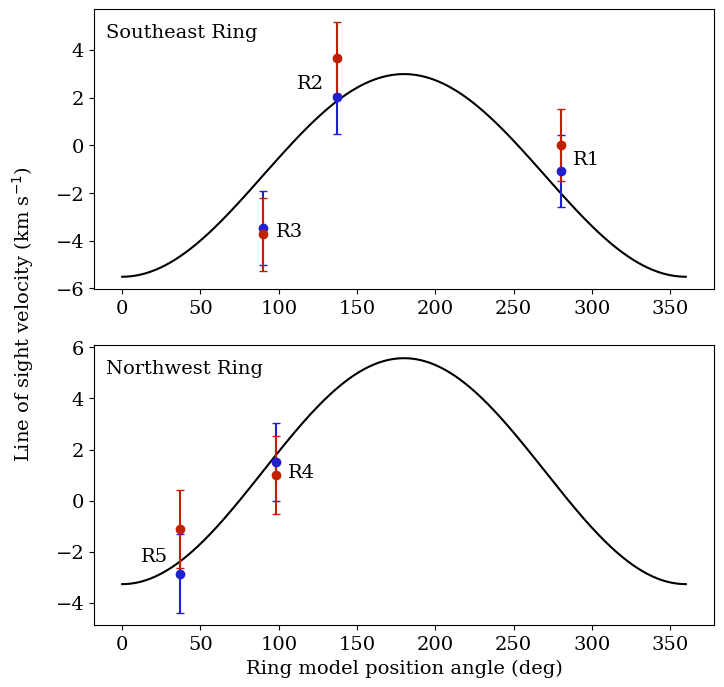}
\caption{Radial velocity at each MRS position measured as described in Figure~\ref{fig:velmod} as determined from [\ion{S}{4}] (blue points) and [\ion{Ne}{3}] (red points). The three points on the SE ring are plotted in the top panel; the two points from the NW ring in the bottom panel. The fits were performed with all five locations simultaneously. The fit is generally consistent with a radial expansion of 5.5~\kms{} from the central source.\label{fig:velexp}}
\end{figure}

Figure~\ref{fig:velexp} shows the velocities from the best-fit expansion model along with the measured velocities. While the fit to the data is imperfect, the model is broadly consistent with radial expansion of $\sim 5.5$~\kms.

\begin{table*}[htbp]
  \caption{Measured line velocities for [\ion{S}{4}] and [\ion{Ne}{3}].\label{tab:vel}}
  \begin{center}
    \newcommand{\mc}[1]{\multicolumn{2}{c}{#1}}
\begin{tabular}{c DDDD}
Species                   & \mc{[\ion{S}{4}]}  & \mc{[\ion{Ne}{3}]} & \mc{[\ion{S}{4}]}  & \mc{[\ion{Ne}{3}]} \\
Wavelength (\microns)     & \mc{10.5105}       & \mc{15.5551}       & \mc{10.5105}       & \mc{15.5551}       \\
Slice Width (km s$^{-1}$) & \mc{37.07}         & \mc{48.18}         & \mc{37.07}         & \mc{48.18}         \\\hline
                          & \mc{Raw}           & \mc{Raw}           & \mc{Corrected}     & \mc{Corrected}     \\
                          & \mc{Velocity}      & \mc{Velocity}      & \mc{Velocity}      & \mc{Velocity}      \\
Pos                       & \mc{(km s$^{-1}$)} & \mc{(km s$^{-1}$)} & \mc{(km s$^{-1}$)} & \mc{(km s$^{-1}$)} \\\hline\hline
\decimals

    CS & -1.0(0.3) & 21.5(0.1) & 2.6(1.5) & 1.8(1.5) \\
EQ & -3.1(0.3) & 22.0(0.1) & 0.5(1.5) & 2.2(1.5) \\
S1 & -4.5(0.2) & 19.5(0.2) & -0.8(1.5) & -0.3(1.5) \\
S2 & -10.4(0.2) & 13.8(0.2) & -6.8(1.5) & -6.0(1.5) \\
R1 & -7.3(0.3) & 17.1(0.2) & -3.7(1.5) & -2.6(1.5) \\
R2 & -9.1(0.3) & 16.0(0.2) & -5.5(1.5) & -3.7(1.5) \\
R3 & -4.2(0.4) & 20.8(0.2) & -0.6(1.6) & 1.0(1.5) \\
R4 & -9.7(0.4) & 13.4(0.3) & -6.1(1.5) & -6.3(1.5) \\
R5 & -4.7(0.3) & 18.1(0.3) & -1.1(1.5) & -1.6(1.5) \\
\end{tabular}

  \end{center}
\end{table*}
\vfill
\subsection{Electron Densities}

\begin{figure}[ht]
\centering\includegraphics[width=\columnwidth]{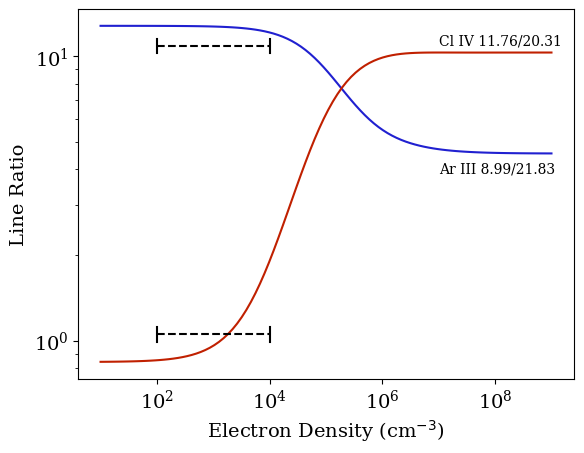}
\caption{The expected line ratios for [\ion{Ar}{3}] (blue) and [\ion{Cl}{4}] (orange). The dashed black line segments indicate the usual range of densities for PN. The [\ion{Ar}{3}] ratio changes very little over the range of densities of interest, while that of [\ion{Cl}{4}] begins to climb steeply.\label{fig:linerat}}
\end{figure}

The wavelength range of the spectra covers two density-sensitive line ratios: [\ion{Ar}{3}] 8.99/21.8~\microns{} and [\ion{Cl}{4}] 11.76/20.31~\microns{}. These are highly complementary density probes with the [\ion{Cl}{4}] ratio sensitive to densities in the range $\sim$10$^{2}$--10$^{6}$~cm$^{-3}$ and [\ion{Ar}{3}] $\sim$10$^{4}$--10$^{7}$~cm$^{-3}$ (Figure~\ref{fig:linerat}). There are no temperature diagnostics in the wavelength range of the observations so in order to use these density diagnostics it is necessary to assume a temperature. Here, we assume the temperature derived by \citet{aller2021} of T$_e = 14000$~K. However, it is important to note that due to the very weak dependence of the line ratios on temperature, there is no significant difference between the results derived for this temperature and those using the canonical temperature usually assumed for PNe (T$_e = 10000$~K). 

\begin{figure}[ht]
\centering\includegraphics[width=\columnwidth]{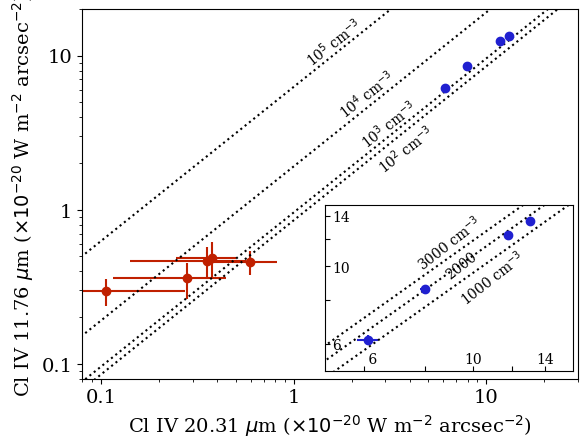}
\caption{The brightness of the [\ion{Cl}{4}] line at 11.76~\microns{} vs the brightness at 20.31~\microns. The blue points represent the central positions, while the red represent the rings. The line ratios associated with densities of 10$^2$ through 10$^5$~cm$^{-3}$ are plotted as dotted lines. The inset focuses on the central position, and densities of 1000, 2000, and 3000~cm$^{-3}$ are plotted. The densities in the central positions are in the 1400-2000~cm$^{-3}$ range; the rings are not detected with enough signal-to-noise to draw meaningful conclusions except perhaps that electron densities significantly  $>10^4$~cm$^{-3}$ are excluded.\label{fig:clrat}}
\end{figure}

Both lines of both ratios are detected with certainty only at the four central positions (CS, EQ, S1, and S2). The measured [\ion{Ar}{3}] ratios (10--30) are outside the range to which the ratio is sensitive (see Figure~\ref{fig:linerat}, blue curve) implying a density $\lesssim$10$^4$~cm$^{-3}$. The [\ion{Cl}{4}] ratios (1.00--1.08) however are more useful and lie in the density sensitive regime  (Figure~\ref{fig:linerat}, orange curve). The central [\ion{Cl}{4}] 11.76~\micron{} vs 20.31~\micron{} brightnesses are plotted in Figure~\ref{fig:clrat} as the blue points, and lie consistently near the n$_e \sim 2000$~cm$^{-3}$ line derived from the line ratio curves in Figure~\ref{fig:linerat}.

To formally derive the densities and their associated uncertainties, we assume Gaussian uncertainties on the line brightness measurements listed in Table~\ref{tab:lines} and perform a Monte Carlo sampling of this distribution to obtain the implied electron density distribution using \texttt{PyNeb} \citep[version 1.1.19, ][]{pyneb}. The line ratios and the derived densities with uncertainties are listed in Table~\ref{tab:dens}.

\begin{table}[htbp]
\caption{Electron densities derived using the [\ion{Cl}{4}] 11.76/20.31~\microns{} ratio.\label{tab:dens}}
\begin{tabular}{c c c}
\hline
Position & Line Ratio & Density \\
 & & (cm$^{-3}$)\\\hline
CS & 1.01  $\pm$ 0.06 & 1430  $\pm$ 510\\
EQ & 1.08  $\pm$ 0.03 & 2010  $\pm$ 260\\
S1 & 1.03  $\pm$ 0.02 & 1630  $\pm$ 190\\
S2 & 1.05  $\pm$ 0.02 & 1790  $\pm$ 140\\
\end{tabular}



\end{table}

The situation in the rings is far less clear, as seen by the red points in Figure~\ref{fig:clrat}. While the 11.76~\micron{} line is detected with some confidence ($\sim 5\sigma$) at all positions, the ``detections'' at 20.31 are typically $1\sigma$ except perhaps at R5; they may simply be upper limits to non-detections that reflect only the noise since a fit was forced at each of these positions. By eye, there are no convincing detections; even at R5, there are neighboring spectral ``features'' of equal magnitude that are likely inadequately corrected fringe effects. If the dual detection at R5 is real, then the density there may be only $\sim 100$~cm$^{-3}$, but if it too is an upper limit, then the non-detections taken together could imply a somewhat higher electron density in the rings than in the inner shell. However, all we can state with any confidence is that the electron density in the rings is unlikely to greatly exceed 10$^4$~cm$^{-3}$.

\subsection{Nature of the Ring Emission\label{sec:ringem}}

To begin the discussion of what the rings are, it is helpful to discuss what they are not. We know that they are not shocks; there is no H$_2$ emission seen at any position in \ngc. We also know that they are not composed of fluorescing PAH molecules; they are also entirely absent in any of the spectra. In fact, we see no evidence for molecules in their gas phase at all at any position of the nebula.

It is possible that atomic line emission (both from neutral hydrogen and highly excited ions) contributes some of the flux seen in the imaging data. To quantify this, we have derived the integrated surface brightness from each of the observed emission lines that are contained in one of the three filter passbands at each spatial position. These are compared against the surface brightnesses measured as the median of a $5 \times 5$ arcsecond box centered on the same location in each filter. The estimated brightness in the filters is not very precise given that we do not know the shape of the full spectrum within each passband since we were unable to reliably detect the continuum in the MRS, but we should be able to draw some broad conclusions. Table~\ref{tab:filterflux} gives the line surface brightness along with the filter surface brightness at each position, omitting the measurements at the CSPN since the emission line data excludes the CSPN while the imager measurements include it.

\begin{table*}[htbp]
\begin{center}
  \caption{Line surface brightness fractions for the three filters\label{tab:filterflux}}
    \newcommand{\mc}[1]{\multicolumn{2}{c}{#1}}
    \newcommand{\mcl}[1]{\multicolumn{2}{c|}{#1}}
    \newcommand{\mcfour}[1]{\multicolumn{4}{c}{#1}}

    \hspace*{-1in}\begin{tabular}{c D | D D@{~}D D@{~}D D@{~}D D@{~}D D}
    \hline
    \multicolumn{23}{c}{F770W Spectral Components}\\
    Position & \mcl{$F_\nu$} & \mc{$F770W$ } & \mcfour{[\ion{Na}{3}]} & \mcfour{\ion{H}{1} 6--5 (Pf $\alpha$)} & \mcfour{\ion{H}{1} 8--6 (Hu $\beta$)} & \mcfour{\ion{H}{1} 10--7}      & \mc{All lines} \\
    & \mcl{(MJy/sr)} &  \multicolumn{20}{c}{($\times 10^{-20}$ W m$^{-2}$ arcsec$^{-2}$)} \\\hline
    \decimals

EQ & 0.847 & 199.5 & 2.0 & (1.0) & 13.4 & (6.7) & 2.7 & (1.4) & 2.8 & (1.4) & (10.5) \\
S1 & 0.586 & 137.9 & 3.1 & (2.2) & 23.3 & (16.9) & 5.4 & (3.9) & 3.0 & (2.2) & (25.2) \\
S2 & 0.513 & 120.9 & 2.3 & (1.9) & 21.9 & (18.1) & 5.5 & (4.6) & 2.5 & (2.1) & (26.6) \\\hline
R1 & 1.300 & 306.2 & 0.0 & (0.0) & 1.1 & (0.4) & 0.1 & (0.0) & 0.6 & (0.2) & (0.6) \\
R2 & 0.518 & 122.0 & 0.1 & (0.1) & 0.9 & (0.8) & 0.3 & (0.3) & 0.2 & (0.2) & (1.3) \\
R3 & 1.451 & 341.7 & 0.3 & (0.1) & 0.5 & (0.1) & 0.6 & (0.2) & 0.5 & (0.1) & (0.5) \\
R4 & 0.746 & 175.6 & 0.4 & (0.2) & 1.0 & (0.6) & 0.1 & (0.1) & 0.9 & (0.5) & (1.4) \\
R5 & 0.405 & 95.3 & 0.7 & (0.7) & 0.5 & (0.5) & 0.3 & (0.3) & 0.1 & (0.1) & (1.7) \\
\end{tabular}
    \par\vspace*{5ex}

    \begin{tabular}{c D | D D@{~}D D@{~}D D@{~}D D@{~}D D@{~}D D}
    \hline
    \multicolumn{27}{c}{F1280W Spectral Components}\\
    Position & \mcl{$F_\nu$} & \mc{$F1280W$ }        &                           \mcfour{[\ion{Cl}{4}]} & \mcfour{\ion{H}{1} 7--6 (Hu $\alpha$)}  & \mcfour{\ion{H}{1} 11--8} & \mcfour{[\ion{Ne}{2}]} & \mcfour{[\ion{Ar}{5}]}   & \mc{All lines} \\
    & \mcl{(MJy/sr)} &  \multicolumn{24}{c}{($\times 10^{-20}$ W m$^{-2}$ arcsec$^{-2}$)} \\\hline
    \decimals

    EQ & 2.537 & 269.0 & 8.6 & (3.2) & 6.0 & (2.2) & 0.4 & (0.2) & 1.4 & (0.5) & 1.7 & (0.6) & (6.8) \\
S1 & 0.958 & 101.6 & 13.5 & (13.3) & 10.3 & (10.1) & 0.9 & (0.9) & 6.5 & (6.4) & 0.3 & (0.3) & (31.0) \\
S2 & 0.819 & 86.8 & 12.4 & (14.2) & 9.4 & (10.8) & 1.0 & (1.2) & 2.7 & (3.1) & 0.1 & (0.1) & (29.4) \\\hline
R1 & 5.154 & 546.7 & 0.5 & (0.1) & 0.5 & (0.1) & 0.0 & (0.0) & 0.0 & (0.0) & 0.0 & (0.0) & (0.2) \\
R2 & 1.762 & 186.9 & 0.5 & (0.3) & 0.5 & (0.3) & 0.0 & (0.0) & 0.1 & (0.1) & 0.2 & (0.1) & (0.7) \\
R3 & 5.294 & 561.5 & 0.4 & (0.1) & 0.5 & (0.1) & 0.1 & (0.0) & 0.1 & (0.0) & 0.1 & (0.0) & (0.2) \\
R4 & 3.219 & 341.4 & 0.3 & (0.1) & 0.4 & (0.1) & 0.0 & (0.0) & 0.0 & (0.0) & 0.1 & (0.0) & (0.2) \\
R5 & 1.391 & 147.6 & 0.5 & (0.3) & 0.7 & (0.5) & 0.0 & (0.0) & 0.1 & (0.1) & 0.2 & (0.1) & (1.0) \\
\end{tabular}
    \par\vspace*{5ex}

    \begin{tabular}{c D| D D@{~}D}
    \hline
    \multicolumn{9}{c}{F2550W Spectral Components}\\
    Position & \mcl{$F_\nu$} & \mc{$F2550W$ } & \mcfour{[\ion{O}{4}]} \\
    & \mcl{(MJy/sr)} &  \multicolumn{6}{c}{($\times 10^{-20}$ W m$^{-2}$ arcsec$^{-2}$)} \\\hline
    \decimals

    EQ & 37.123 & 1492.2 & 1469.2 & (98.5) \\
S1 & 11.051 & 444.2 & 492.1 & (110.8) \\
S2 & 10.701 & 430.1 & 223.6 & (52.0) \\\hline
R1 & 32.979 & 1325.6 & 6.0 & (0.5) \\
R2 & 8.393 & 337.3 & 4.8 & (1.4) \\
R3 & 31.609 & 1270.5 & 9.9 & (0.8) \\
R4 & 21.054 & 846.3 & 4.8 & (0.6) \\
R5 & 8.188 & 329.1 & 3.2 & (1.0) \\
\end{tabular}

\end{center}
    \tablecomments{the surface brightness for each line is expressed in physical units ($\times 10^{-20}$ W m$^{-2}$ arcsec$^{-2}$) as well as its percentage with respect to the filter surface brightness (in parentheses). The final ``All Lines'' column gives the total percentage of all lines with respect to the filter surface brightness at that position. Since the F2550W image contains only the [\ion{O}{4}] line, no summary column is given.}
\end{table*}

The table shows two results clearly: first, even though we have avoided the extremely bright [\ion{S}{4}] and [\ion{Ne}{3}] lines with our choice of filters, approximately 25--30\% of the emission from the inner shell in the F770W and F1280W bands comes from line emission, predominantly the hydrogen Pfund $\alpha$ and Humphreys $\beta$ lines in F770W, while F1280W has emission from [\ion{Cl}{4}] and hydrogen Humphreys $\alpha$, with a lesser contribution from [\ion{Ne}{2}]. Even greater, essentially \emph{all} of the F2550W emission from the inner shell comes from [\ion{O}{4}]. 

Second, and conversely, almost none of the emission from the rings is contributed by line emission. Less than 1\% of the ring flux in F770W and F1280W can be attributed to the sum of all lines within the passband. Even in F2550W, the overwhelmingly bright [\ion{O}{4}] line contributes no more than 1.5\% of the flux in any part of the rings. The two locations where the contribution is $\gtrsim 1$\% also happen to be where the rings are faintest and may overlap the fringes of the inner shell, thus leading to a partially contaminated measurement. Even so, we conclude that essentially none of the flux from the rings seen in the MIRI images comes from line emission.

Therefore, for the three MIRI filters chosen here, the rings must shine entirely by thermal dust emission. Only in two MIRI filters that we did not select, F1000W and F1500W (and F1550C if it could be used for imaging), would we expect to see line emission in the rings--- from [\ion{S}{4}] and [\ion{Ne}{3}], respectively---all others will be dominated by dust emission.

\subsection{Dust Color Temperature}
\begin{figure}[ht]
\centering\includegraphics[width=\columnwidth]{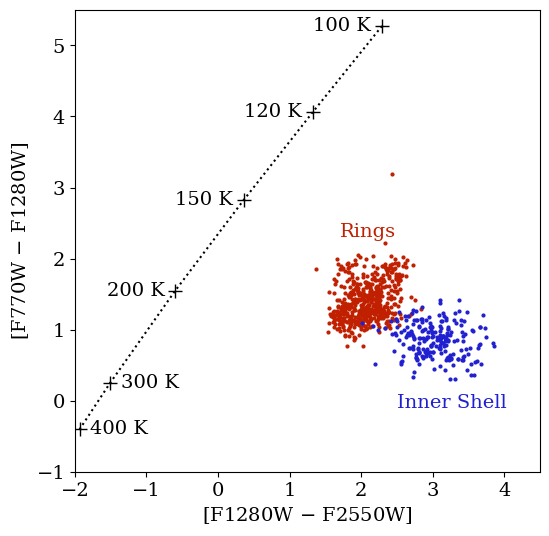}
\caption{Color-color diagram of the rings vs the inner shell. The images were spatially rebinned to $5\times 5$~arcsecond squares, then only those squares for which the mean surface brightness in F1280W was greater than 0.8 MJy/sr were plotted. Values from within the rings were plotted as red and from within the inner shell as blue. The portion of the inner shell where the rings overlap the shell was masked to reduce confusion.\label{fig:colors}}
\end{figure}

Since the rings appear to be purely dust emission, we attempt constructing color-color maps to derive a characteristic color temperature of the ring material. Figure~\ref{fig:colors} shows surface brightness ratios for positions within both the rings and the inner shell in a [F770W -- F1280W] vs [F1280W -- F2550W] color-color diagram. The colors of a pure blackbody are also plotted. The relatively compact set of F770W/F1280W ratios for the rings implies a relatively uniform color temperature of $\sim$~200~K, while the F1280W/F2550W ratio implies a relatively uniform $\sim$~110~K. This range of temperatures are quite common for  dusty PN \citep{zhang1993,stang2012,bilikova2012}. It is not surprising that a single color temperature does not apply to the two filter pairs \citep[\eg][]{otsuka2017} since there is certainly a continuum of grain sizes being heated to different temperatures.  The inner shell is distinct from the rings given that the [\ion{O}{4}] emission causes the shell to be much brighter at F2550W, thus moving the shell locus to the right (redder side) of the diagram. These temperatures are in broad agreement with the $\sim180$~K found by R10 using only the WISE fluxes given that the ring flux in the W3 filter is ``contaminated'' by the [\ion{S}{4}] and [\ion{Ne}{3}] lines. However, even in the WISE data it was clear that multiple temperatures were needed to fit the SEDs.

Since the rings are large ($\sim 0.19$~pc from the CSPN), simple equilibrium heating of large grains will not produce these temperatures. Stochastic heating of very small grains can achieve these temperatures at these distances, but the grains cannot have a significant PAH component since their emission is not present. \citet{bsalas2009} found an anti-correlation in their Spitzer survey of PN in the LMC and SMC between the presence of [\ion{O}{4}] emission and PAH emission. Perhaps the UV field is strong enough to destroy PAHs or prevent their formation, leaving only graphitic grains (or some other mineral) present to be heated.

\subsection{Formation of the Rings}

Unfortunately, our observational results, taken by themselves, do not offer any direct insight into the formation of the rings; we have said something about what they are, but cannot say how they got there. However, there is some modeling work by other groups that may be applicable. In particular, Model A4 of \citet{sg2018} bears some similarity to \ngc, in particular the tilted view in their Figure~8 (third row, right column). This model investigates a binary star system with a common envelope where there is a significant thermal pulse which produces large changes in the density of material into which the PN is expanding. Applied to \ngc{}, the concept would be that a period of heavy mass loss formed the dense regions that would become the source material for the rings, followed by fast jets/winds that carved out the material along the poles to yield a more ring-like structure.

While \citet{sg2018}'s model image is intriguing, we still have a number of hurdles to consider before assigning \ngc{} to this case. The rings in \ngc{} are located within the outer shell, while the model contains them within the inner shell. The large bubbles in the model (out to $\pm 0.2$~pc) may be present in \ngc, but are too faint to observe, perhaps given the older age of \ngc{} \citep[$\sim4000$ yrs,][]{aller2021}, though the material outside the rings in Figure~\ref{fig:ejecta} may be the vestiges of such bubbles as they interacted with the ring material. The model also give little information about the properties and separation of the stars; those in \ngc{} are hotter and more massive than those considered in the model, and the 9-yr-period/6-AU separation is certain larger than those modeled. Finally, and most importantly, \citet{sg2018} do not consider dust in their models, while the rings in \ngc{} are observed primarily by their cold dust emission. Still, tuning such a model to \ngc{}'s properties might bear a better resemblance.

The work of \citet{akashi2018} also contains some interesting parallels. In their models they assume a large, slow, mostly spherical, mass ejection history. On top of this, however, there is a low velocity but high mass loss rate ejection from the AGB progenitor triggered by binary interactions. Finally two higher-velocity jets are injected into the wind/shell. 

Their models R3 (and R4), where the shell is large and the jets launch continuously, but without (and with, respectively) radiative cooling in an optically thin gas, plausibly resemble \ngc{} in the projected images shown in their Figure~11 (third and fourth rows). The chief problem comparing image here is the fact that their models explored ages only one tenth of that of \ngc, though the other difficulties mentioned for the models of \citet{sg2018} are relevant here.

\section{Conclusion} \label{sec:concl}

We have presented mid-infrared imaging and spectroscopy of \ngc{}, focusing on the distinctive pair of rings that appear only at infrared wavelengths. The imaging at 7.7, 12.8, and 25.5~\microns{} reveals a wealth of turbulent structure in the rings, while still showing them to be relatively cohesive structures. The spectroscopy is dominated by emission from the [\ion{S}{4}], [\ion{Ne}{3}], and [\ion{O}{4}] at 10.511, 15.555, and 25.890~\microns, respectively; all other lines are nearly two orders of magnitude fainter except for [\ion{Ar}{3}] at 8.991~\microns{} and [\ion{S}{3}] at 18.713~\microns) which are one and a half orders of magnitude fainter.

While atomic lines account for a quarter to a third of the emission seen from the inner shell at F770W and F1280W, the [\ion{O}{4}] line accounts for almost all of the inner shell emission at F2550W. On the other hand, the spectra show that there is almost no line emission from the rings themselves. Given the lack of PAH or H$_2$ emission anywhere in the nebula, the ring emission must come from stochastically heated graphitic or other non-PAH bearing dust grains.

While our observations do not yield direct information about how the rings formed, the slow doppler velocities in the rings suggest that they were formed from material ejected in a slow, heavy mass loss phase from the PN progenitor, with further shaping at later times from faster winds that created the rest of the optically visible nebula. Regardless, the new data do complete the picture of the rings being cool dusty structures embedded in the tenuous outer shell of a very complex, but fascinating planetary nebula.
\newpage
\vfill\null

\begin{acknowledgments}
This work is primarily based on observations made with the NASA/ESA/CSA James Webb Space Telescope. The data were retrieved from the Mikulski Archive for Space Telescopes at the Space Telescope Science Institute, which is operated by the Association of Universities for Research in Astronomy, Inc., under NASA contract NAS 5-03127 for JWST. These observations are associated with program \#1238 and the raw data (``uncal'') files that were processed here may be found at \dataset[https://dx.doi.org/10.17909/za5q-ss90]{https://dx.doi.org/10.17909/za5q-ss90}.

Ancillary observations were obtained at the Hale Telescope, Palomar Observatory, as part of a collaborative agreement between the Caltech Optical Observatories and the Jet Propulsion Laboratory [operated by Caltech for NASA].

The work of MER and KW was carried out at the Jet Propulsion Laboratory, California Institute of Technology, under a contract with the National Aeronautics and Space Administration.

AA acknowledges financial support from the Spanish Virtual Observatory project funded by the Spanish Ministry of Science and Innovation/State Agency of Research MCIN/AEI/10.13039/501100011033 through grant PID2020112949GB-I00. AA also acknowledges support from the I+D+i project PID2019-105203GB-C21 founded
by Spanish AEI (MICIU) grant 10.13039/501100011033. 


DJ acknowledges support from the Agencia Estatal de Investigaci\'on del Ministerio de Ciencia, Innovaci\'on y Universidades (MCIU/AEI) and the European Regional Development Fund (ERDF) with reference PID-2022-136653NA-I00 (DOI:10.13039/501100011033). DJ also acknowledges support from the Agencia Estatal de Investigaci\'on del Ministerio de Ciencia, Innovaci\'on y Universidades (MCIU/AEI) and the the European Union NextGenerationEU/PRTR with reference CNS2023-143910 (DOI:10.13039/501100011033).

LFM acknowledges support from grants PID2020-114461GB-I00, \mbox{PID2023-146295NB-I00}, and CEX2021-001131-S, funded by MCIN/AEI/10.13039/501100011033.

\end{acknowledgments}

\vspace{5mm}
\facilities{JWST(MIRI), WISE, Palomar:5m(WIRC)}

\software{Scipy \citep{scipy2020}, Astropy \citep{2013A&A...558A..33A,2018AJ....156..123A},
JWST pipeline \citep{bushouse2022}, ICORE \citep{masci2009}, PyNeb \citep{pyneb}}

\newpage

\appendix

\section{Palomar Planning Image\label{app:palo}}
Near-infrared images of \ngc{} were taken with the WIRC instrument \citep{wirc,wircpol} on the Palomar 5-m telescope on 2019 November 4 UTC to aid the JWST observation planning. Data were taken with the $H$ (1.64~\microns) and $K_s$ (2.15~\microns) filters, but the sky quality was somewhat variable and the $H$ filter data were not of sufficient quality nor quantity to aid in the planning.

\begin{figure}[ht]
\includegraphics[width=\columnwidth]{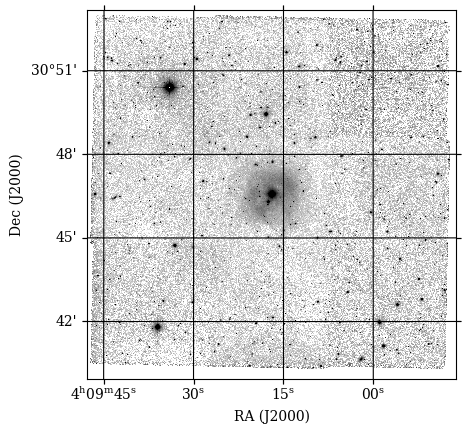}
\includegraphics[width=\columnwidth]{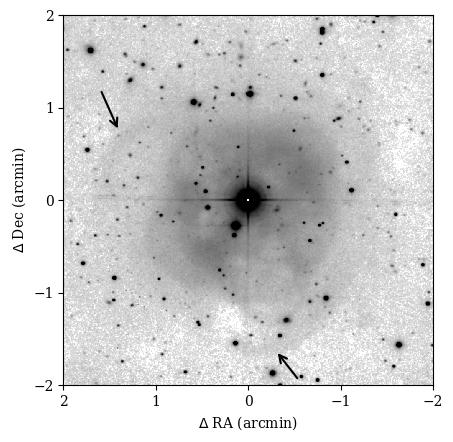}
\caption{Upper: full $K_s$ image obtained with the WIRC instrument at the Palomar 5-m telescope. Lower: inner 4 arcminutes of the image, lightly smoothed with a gaussian kernel to highlight the nebulosity; the location of the south-east ring is indicated by arrows.\label{fig:ks}}
\end{figure}

For the $K_s$ filter, there were 60 exposures in total, each 24 integrations of 5 seconds each. Only the best 40 of the exposures were included in the final coadd; 20 were rejected due to bad seeing or non-uniform backgrounds. The data were background-subtracted and flat-fielded; bad pixels and columns were masked, as were several ghosts of the \ngc{} central source. The images were coadded by aligning on the central source and stacking, producing a final image with 4800 seconds total exposure time on the nebula (Figure~\ref{fig:ks}). The telescope was ``dithered'' in a round-robin fashion; the nebula was placed in the center of one quadrant, starting in the upper left, then moved in a  clockwise direction through all four quadrants. Residuals in this process give the final image a slight $3\times3$  tile appearance---each tile being $\sim$ 4 arcminutes in extent---with the central tile receiving the full exposure time, the 4 middle edge tiles receiving half, and the 4 corner tiles receiving only a quarter of the exposure time. The central tile encloses the full nebula, so all areas of interest for planning are at full sensitivity.

The image reveals a number of background stars and galaxies upon which the nebula is superimposed. There is possibly even a galaxy cluster on the north-northeastern side of the nebula. Though faint, this is the shortest wavelength image where the rings are unambiguously detected. The $H$ images were also stacked, but the rings are not seen, with the caveat that the final usable exposure time was only 1080 seconds.

\section{Rest Velocity Correction\label{app:velo}}

For each of the inner shell positions, the ensemble velocity of the detected lines was estimated by taking the median of the velocity values of all lines that were detected with a SNR of at least 7; if $v_{line, pos}$ is the velocity of each line at each position (20 lines at 4 positions for determining the correction), we can represent this as

$$ \tilde{v}_{pos} = \mathrm{median}(v_{line_1, pos} \ldots\, v_{line_n, pos}), \mathrm{~for~SNR > 7}$$

\begin{figure}[ht]
  \includegraphics[width=\columnwidth]{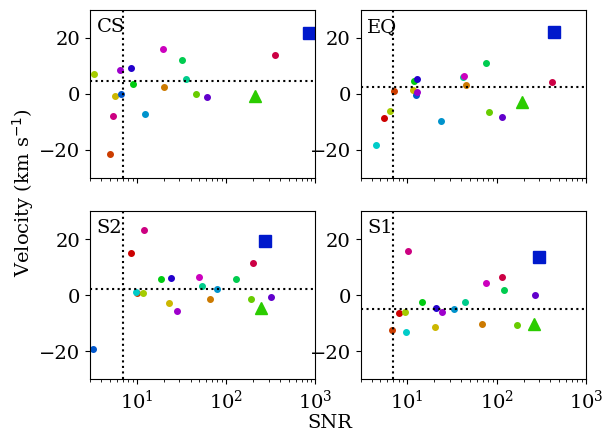}
    \includegraphics[width=\columnwidth]{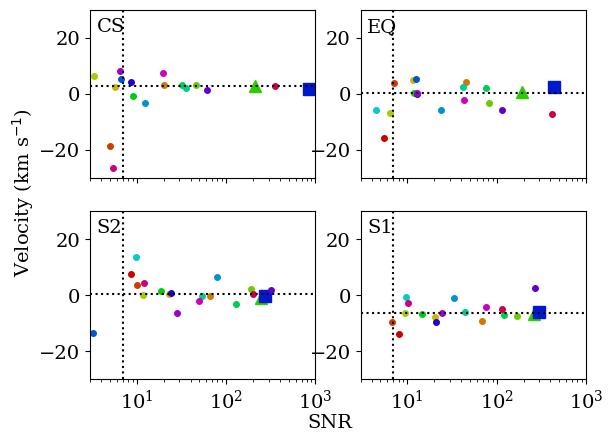}
\caption{Upper: velocities as measured directly from the line profile fits. Lower: velocities after correction. The horizontal dotted lines represent the median of each distribution. The vertical dotted lines show the SNR = 7 cutoff used for selecting the sample used in the median. The larger green triangular and blue square symbols represent [\ion{S}{4}] and [\ion{Ne}{3}], respectively. These lines are the only ones bright enough for estimating velocities in the rings, so having good corrections for them is especially critical. \label{fig:velfix}}
\end{figure}

This threshold allowed between 12 and 19 lines to be used in each estimate. The raw velocity values are presented in Figure~\ref{fig:velfix}, top; the median velocities are plotted as the horizontal dotted lines. We then subtracted these values from each line velocity, then calculated the median difference across the four positions to estimate the velocity correction for each line:

$$ \Delta v_{line} = \mathrm{median}(v_{line, pos_1} - \tilde{v}_{pos_1} \ldots\, v_{line, pos_n} - \tilde{v}_{pos_n})$$

This $\Delta V_{line}$ was then subtracted from every velocity measurement to obtain the values used in Table~\ref{tab:vel} and Figure~\ref{fig:velexp}. The results of this subtraction are presented in the right hand plot of Figure~\ref{fig:velfix}; the reduction in the deviation is significant, especially for the high SNR lines. The median-absolute-deviation of the corrected velocities from their median is approximately 1.5~\kms{} and we carry this forward as the uncertainly in the corrected velocity values.

\clearpage
\bibliography{ngc1514}{}

\begin{thebibliography}{}
\expandafter\ifx\csname natexlab\endcsname\relax\def\natexlab#1{#1}\fi
\providecommand{\url}[1]{\href{#1}{#1}}
\providecommand{\dodoi}[1]{doi:~\href{http://doi.org/#1}{\nolinkurl{#1}}}
\providecommand{\doeprint}[1]{\href{http://ascl.net/#1}{\nolinkurl{http://ascl.net/#1}}}
\providecommand{\doarXiv}[1]{\href{https://arxiv.org/abs/#1}{\nolinkurl{https://arxiv.org/abs/#1}}}

\bibitem[{{Akashi} {et~al.}(2018){Akashi}, {Bear}, \& {Soker}}]{akashi2018}
{Akashi}, M., {Bear}, E., \& {Soker}, N. 2018, \mnras, 475, 4794,
  \dodoi{10.1093/mnras/sty029}

\bibitem[{{Akras} {et~al.}(2016){Akras}, {Clyne}, {Boumis}, {Monteiro},
  {Gon{\c{c}}alves}, {Redman}, \& {Williams}}]{akras2016}
{Akras}, S., {Clyne}, N., {Boumis}, P., {et~al.} 2016, \mnras, 457, 3409,
  \dodoi{10.1093/mnras/stw038}

\bibitem[{{Aller} {et~al.}(2015){Aller}, {Montesinos}, {Miranda}, {Solano}, \&
  {Ulla}}]{aller2015}
{Aller}, A., {Montesinos}, B., {Miranda}, L.~F., {Solano}, E., \& {Ulla}, A.
  2015, \mnras, 448, 2822, \dodoi{10.1093/mnras/stv196}

\bibitem[{{Aller} {et~al.}(2021){Aller}, {V{\'a}zquez}, {Olgu{\'\i}n},
  {Miranda}, \& {Ressler}}]{aller2021}
{Aller}, A., {V{\'a}zquez}, R., {Olgu{\'\i}n}, L., {Miranda}, L.~F., \&
  {Ressler}, M.~E. 2021, \mnras, 504, 4806, \dodoi{10.1093/mnras/stab1233}

\bibitem[{{Astropy Collaboration} {et~al.}(2013){Astropy Collaboration},
  {Robitaille}, {Tollerud}, {Greenfield}, {Droettboom}, {Bray}, {Aldcroft},
  {Davis}, {Ginsburg}, {Price-Whelan}, {Kerzendorf}, {Conley}, {Crighton},
  {Barbary}, {Muna}, {Ferguson}, {Grollier}, {Parikh}, {Nair}, {Unther},
  {Deil}, {Woillez}, {Conseil}, {Kramer}, {Turner}, {Singer}, {Fox}, {Weaver},
  {Zabalza}, {Edwards}, {Azalee Bostroem}, {Burke}, {Casey}, {Crawford},
  {Dencheva}, {Ely}, {Jenness}, {Labrie}, {Lim}, {Pierfederici}, {Pontzen},
  {Ptak}, {Refsdal}, {Servillat}, \& {Streicher}}]{2013A&A...558A..33A}
{Astropy Collaboration}, {Robitaille}, T.~P., {Tollerud}, E.~J., {et~al.} 2013,
  \aap, 558, A33, \dodoi{10.1051/0004-6361/201322068}

\bibitem[{{Astropy Collaboration} {et~al.}(2018){Astropy Collaboration},
  {Price-Whelan}, {Sip{\H{o}}cz}, {G{\"u}nther}, {Lim}, {Crawford}, {Conseil},
  {Shupe}, {Craig}, {Dencheva}, {Ginsburg}, {VanderPlas}, {Bradley},
  {P{\'e}rez-Su{\'a}rez}, {de Val-Borro}, {Aldcroft}, {Cruz}, {Robitaille},
  {Tollerud}, {Ardelean}, {Babej}, {Bach}, {Bachetti}, {Bakanov}, {Bamford},
  {Barentsen}, {Barmby}, {Baumbach}, {Berry}, {Biscani}, {Boquien}, {Bostroem},
  {Bouma}, {Brammer}, {Bray}, {Breytenbach}, {Buddelmeijer}, {Burke},
  {Calderone}, {Cano Rodr{\'\i}guez}, {Cara}, {Cardoso}, {Cheedella}, {Copin},
  {Corrales}, {Crichton}, {D'Avella}, {Deil}, {Depagne}, {Dietrich}, {Donath},
  {Droettboom}, {Earl}, {Erben}, {Fabbro}, {Ferreira}, {Finethy}, {Fox},
  {Garrison}, {Gibbons}, {Goldstein}, {Gommers}, {Greco}, {Greenfield},
  {Groener}, {Grollier}, {Hagen}, {Hirst}, {Homeier}, {Horton}, {Hosseinzadeh},
  {Hu}, {Hunkeler}, {Ivezi{\'c}}, {Jain}, {Jenness}, {Kanarek}, {Kendrew},
  {Kern}, {Kerzendorf}, {Khvalko}, {King}, {Kirkby}, {Kulkarni}, {Kumar},
  {Lee}, {Lenz}, {Littlefair}, {Ma}, {Macleod}, {Mastropietro}, {McCully},
  {Montagnac}, {Morris}, {Mueller}, {Mumford}, {Muna}, {Murphy}, {Nelson},
  {Nguyen}, {Ninan}, {N{\"o}the}, {Ogaz}, {Oh}, {Parejko}, {Parley}, {Pascual},
  {Patil}, {Patil}, {Plunkett}, {Prochaska}, {Rastogi}, {Reddy Janga},
  {Sabater}, {Sakurikar}, {Seifert}, {Sherbert}, {Sherwood-Taylor}, {Shih},
  {Sick}, {Silbiger}, {Singanamalla}, {Singer}, {Sladen}, {Sooley},
  {Sornarajah}, {Streicher}, {Teuben}, {Thomas}, {Tremblay}, {Turner},
  {Terr{\'o}n}, {van Kerkwijk}, {de la Vega}, {Watkins}, {Weaver}, {Whitmore},
  {Woillez}, {Zabalza}, \& {Astropy Contributors}}]{2018AJ....156..123A}
{Astropy Collaboration}, {Price-Whelan}, A.~M., {Sip{\H{o}}cz}, B.~M., {et~al.}
  2018, \aj, 156, 123, \dodoi{10.3847/1538-3881/aabc4f}

\bibitem[{Balick(2004)}]{balick2004}
Balick, B. 2004, The Astronomical Journal, 127, 2262, \dodoi{10.1086/382518}

\bibitem[{Bernard-Salas {et~al.}(2009)Bernard-Salas, Peeters, Sloan,
  Gutenkunst, Matsuura, Tielens, Zijlstra, \& Houck}]{bsalas2009}
Bernard-Salas, J., Peeters, E., Sloan, G.~C., {et~al.} 2009, The Astrophysical
  Journal, 699, 1541, \dodoi{10.1088/0004-637X/699/2/1541}

\bibitem[{Bilíková {et~al.}(2012)Bilíková, Chu, Gruendl, Su, \&
  Marco}]{bilikova2012}
Bilíková, J., Chu, Y.-H., Gruendl, R.~A., Su, K. Y.~L., \& Marco, O.~D. 2012,
  The Astrophysical Journal Supplement Series, 200, 3,
  \dodoi{10.1088/0067-0049/200/1/3}

\bibitem[{{Boffin} \& {Jones}(2019)}]{boffin2019}
{Boffin}, H. M.~J., \& {Jones}, D. 2019, {The Importance of Binaries in the
  Formation and Evolution of Planetary Nebulae} (Springer Cham),
  \dodoi{10.1007/978-3-030-25059-1}

\bibitem[{{Brown} {et~al.}(2014){Brown}, {Jarrett}, \& {Cluver}}]{brown2014}
{Brown}, M.~J.~I., {Jarrett}, T.~H., \& {Cluver}, M.~E. 2014, \pasa, 31, e049,
  \dodoi{10.1017/pasa.2014.44}

\bibitem[{{Bushouse} {et~al.}(2022){Bushouse}, {Eisenhamer}, {Dencheva},
  {Davies}, {Greenfield}, {Morrison}, {Hodge}, {Simon}, {Grumm}, {Droettboom},
  {Slavich}, {Sosey}, {Pauly}, {Miller}, {Jedrzejewski}, {Hack}, {Davis},
  {Crawford}, {Law}, {Gordon}, {Regan}, {Cara}, {MacDonald}, {Bradley},
  {Shanahan}, {Jamieson}, {Teodoro}, \& {Williams}}]{bushouse2022}
{Bushouse}, H., {Eisenhamer}, J., {Dencheva}, N., {et~al.} 2022, {JWST
  Calibration Pipeline}, 1.6.2,  Zenodo, \dodoi{10.5281/zenodo.7041998}

\bibitem[{{Chu} {et~al.}(1987){Chu}, {Jacoby}, \& {Arendt}}]{chu1987}
{Chu}, Y.-H., {Jacoby}, G.~H., \& {Arendt}, R. 1987, \apjs, 64, 529,
  \dodoi{10.1086/191207}

\bibitem[{{ESA}(1997)}]{hip1997}
{ESA}. 1997, {The Hipparcos and Tycho Catalogues}, ESA SP-1200

\bibitem[{{Gaia Collaboration} {et~al.}(2023){Gaia Collaboration}, {Vallenari},
  {Brown}, {Prusti}, {de Bruijne}, {Arenou}, {Babusiaux}, {Biermann},
  {Creevey}, {Ducourant}, {Evans}, {Eyer}, {Guerra}, {Hutton}, {Jordi},
  {Klioner}, {Lammers}, {Lindegren}, {Luri}, {Mignard}, {Panem}, {Pourbaix},
  {Randich}, {Sartoretti}, {Soubiran}, {Tanga}, {Walton}, {Bailer-Jones},
  {Bastian}, {Drimmel}, {Jansen}, {Katz}, {Lattanzi}, {van Leeuwen}, {Bakker},
  {Cacciari}, {Casta{\~n}eda}, {De Angeli}, {Fabricius}, {Fouesneau},
  {Fr{\'e}mat}, {Galluccio}, {Guerrier}, {Heiter}, {Masana}, {Messineo},
  {Mowlavi}, {Nicolas}, {Nienartowicz}, {Pailler}, {Panuzzo}, {Riclet}, {Roux},
  {Seabroke}, {Sordo}, {Th{\'e}venin}, {Gracia-Abril}, {Portell}, {Teyssier},
  {Altmann}, {Andrae}, {Audard}, {Bellas-Velidis}, {Benson}, {Berthier},
  {Blomme}, {Burgess}, {Busonero}, {Busso}, {C{\'a}novas}, {Carry}, {Cellino},
  {Cheek}, {Clementini}, {Damerdji}, {Davidson}, {de Teodoro}, {Nu{\~n}ez
  Campos}, {Delchambre}, {Dell'Oro}, {Esquej}, {Fern{\'a}ndez-Hern{\'a}ndez},
  {Fraile}, {Garabato}, {Garc{\'\i}a-Lario}, {Gosset}, {Haigron}, {Halbwachs},
  {Hambly}, {Harrison}, {Hern{\'a}ndez}, {Hestroffer}, {Hodgkin}, {Holl},
  {Jan{\ss}en}, {Jevardat de Fombelle}, {Jordan}, {Krone-Martins}, {Lanzafame},
  {L{\"o}ffler}, {Marchal}, {Marrese}, {Moitinho}, {Muinonen}, {Osborne},
  {Pancino}, {Pauwels}, {Recio-Blanco}, {Reyl{\'e}}, {Riello}, {Rimoldini},
  {Roegiers}, {Rybizki}, {Sarro}, {Siopis}, {Smith}, {Sozzetti}, {Utrilla},
  {van Leeuwen}, {Abbas}, {{\'A}brah{\'a}m}, {Abreu Aramburu}, {Aerts},
  {Aguado}, {Ajaj}, {Aldea-Montero}, {Altavilla}, {{\'A}lvarez}, {Alves},
  {Anders}, {Anderson}, {Anglada Varela}, {Antoja}, {Baines}, {Baker},
  {Balaguer-N{\'u}{\~n}ez}, {Balbinot}, {Balog}, {Barache}, {Barbato},
  {Barros}, {Barstow}, {Bartolom{\'e}}, {Bassilana}, {Bauchet}, {Becciani},
  {Bellazzini}, {Berihuete}, {Bernet}, {Bertone}, {Bianchi}, {Binnenfeld},
  {Blanco-Cuaresma}, {Blazere}, {Boch}, {Bombrun}, {Bossini}, {Bouquillon},
  {Bragaglia}, {Bramante}, {Breedt}, {Bressan}, {Brouillet}, {Brugaletta},
  {Bucciarelli}, {Burlacu}, {Butkevich}, {Buzzi}, {Caffau}, {Cancelliere},
  {Cantat-Gaudin}, {Carballo}, {Carlucci}, {Carnerero}, {Carrasco},
  {Casamiquela}, {Castellani}, {Castro-Ginard}, {Chaoul}, {Charlot}, {Chemin},
  {Chiaramida}, {Chiavassa}, {Chornay}, {Comoretto}, {Contursi}, {Cooper},
  {Cornez}, {Cowell}, {Crifo}, {Cropper}, {Crosta}, {Crowley}, {Dafonte},
  {Dapergolas}, {David}, {David}, {de Laverny}, {De Luise}, {De March}, {De
  Ridder}, {de Souza}, {de Torres}, {del Peloso}, {del Pozo}, {Delbo},
  {Delgado}, {Delisle}, {Demouchy}, {Dharmawardena}, {Di Matteo}, {Diakite},
  {Diener}, {Distefano}, {Dolding}, {Edvardsson}, {Enke}, {Fabre}, {Fabrizio},
  {Faigler}, {Fedorets}, {Fernique}, {Fienga}, {Figueras}, {Fournier},
  {Fouron}, {Fragkoudi}, {Gai}, {Garcia-Gutierrez}, {Garcia-Reinaldos},
  {Garc{\'\i}a-Torres}, {Garofalo}, {Gavel}, {Gavras}, {Gerlach}, {Geyer},
  {Giacobbe}, {Gilmore}, {Girona}, {Giuffrida}, {Gomel}, {Gomez},
  {Gonz{\'a}lez-N{\'u}{\~n}ez}, {Gonz{\'a}lez-Santamar{\'\i}a},
  {Gonz{\'a}lez-Vidal}, {Granvik}, {Guillout}, {Guiraud},
  {Guti{\'e}rrez-S{\'a}nchez}, {Guy}, {Hatzidimitriou}, {Hauser}, {Haywood},
  {Helmer}, {Helmi}, {Sarmiento}, {Hidalgo}, {Hilger}, {H{\l}adczuk}, {Hobbs},
  {Holland}, {Huckle}, {Jardine}, {Jasniewicz}, {Jean-Antoine Piccolo},
  {Jim{\'e}nez-Arranz}, {Jorissen}, {Juaristi Campillo}, {Julbe}, {Karbevska},
  {Kervella}, {Khanna}, {Kontizas}, {Kordopatis}, {Korn}, {K{\'o}sp{\'a}l},
  {Kostrzewa-Rutkowska}, {Kruszy{\'n}ska}, {Kun}, {Laizeau}, {Lambert},
  {Lanza}, {Lasne}, {Le Campion}, {Lebreton}, {Lebzelter}, {Leccia}, {Leclerc},
  {Lecoeur-Taibi}, {Liao}, {Licata}, {Lindstr{\o}m}, {Lister}, {Livanou},
  {Lobel}, {Lorca}, {Loup}, {Madrero Pardo}, {Magdaleno Romeo}, {Managau},
  {Mann}, {Manteiga}, {Marchant}, {Marconi}, {Marcos}, {Marcos Santos},
  {Mar{\'\i}n Pina}, {Marinoni}, {Marocco}, {Marshall}, {Martin Polo},
  {Mart{\'\i}n-Fleitas}, {Marton}, {Mary}, {Masip}, {Massari},
  {Mastrobuono-Battisti}, {Mazeh}, {McMillan}, {Messina}, {Michalik}, {Millar},
  {Mints}, {Molina}, {Molinaro}, {Moln{\'a}r}, {Monari}, {Mongui{\'o}},
  {Montegriffo}, {Montero}, {Mor}, {Mora}, {Morbidelli}, {Morel}, {Morris},
  {Muraveva}, {Murphy}, {Musella}, {Nagy}, {Noval}, {Oca{\~n}a}, {Ogden},
  {Ordenovic}, {Osinde}, {Pagani}, {Pagano}, {Palaversa}, {Palicio},
  {Pallas-Quintela}, {Panahi}, {Payne-Wardenaar}, {Pe{\~n}alosa Esteller},
  {Penttil{\"a}}, {Pichon}, {Piersimoni}, {Pineau}, {Plachy}, {Plum}, {Poggio},
  {Pr{\v{s}}a}, {Pulone}, {Racero}, {Ragaini}, {Rainer}, {Raiteri}, {Rambaux},
  {Ramos}, {Ramos-Lerate}, {Re Fiorentin}, {Regibo}, {Richards}, {Rios Diaz},
  {Ripepi}, {Riva}, {Rix}, {Rixon}, {Robichon}, {Robin}, {Robin}, {Roelens},
  {Rogues}, {Rohrbasser}, {Romero-G{\'o}mez}, {Rowell}, {Royer}, {Ruz Mieres},
  {Rybicki}, {Sadowski}, {S{\'a}ez N{\'u}{\~n}ez}, {Sagrist{\`a} Sell{\'e}s},
  {Sahlmann}, {Salguero}, {Samaras}, {Sanchez Gimenez}, {Sanna},
  {Santove{\~n}a}, {Sarasso}, {Schultheis}, {Sciacca}, {Segol}, {Segovia},
  {S{\'e}gransan}, {Semeux}, {Shahaf}, {Siddiqui}, {Siebert}, {Siltala},
  {Silvelo}, {Slezak}, {Slezak}, {Smart}, {Snaith}, {Solano}, {Solitro},
  {Souami}, {Souchay}, {Spagna}, {Spina}, {Spoto}, {Steele},
  {Steidelm{\"u}ller}, {Stephenson}, {S{\"u}veges}, {Surdej}, {Szabados},
  {Szegedi-Elek}, {Taris}, {Taylor}, {Teixeira}, {Tolomei}, {Tonello}, {Torra},
  {Torra}, {Torralba Elipe}, {Trabucchi}, {Tsounis}, {Turon}, {Ulla}, {Unger},
  {Vaillant}, {van Dillen}, {van Reeven}, {Vanel}, {Vecchiato}, {Viala},
  {Vicente}, {Voutsinas}, {Weiler}, {Wevers}, {Wyrzykowski}, {Yoldas}, {Yvard},
  {Zhao}, {Zorec}, {Zucker}, \& {Zwitter}}]{gaiadr3}
{Gaia Collaboration}, {Vallenari}, A., {Brown}, A.~G.~A., {et~al.} 2023, \aap,
  674, A1, \dodoi{10.1051/0004-6361/202243940}

\bibitem[{{Garc{\'\i}a-Segura} {et~al.}(2018){Garc{\'\i}a-Segura}, {Ricker}, \&
  {Taam}}]{sg2018}
{Garc{\'\i}a-Segura}, G., {Ricker}, P.~M., \& {Taam}, R.~E. 2018, \apj, 860,
  19, \dodoi{10.3847/1538-4357/aac08c}

\bibitem[{{Jones} \& {Boffin}(2017)}]{jones2017b}
{Jones}, D., \& {Boffin}, H. M.~J. 2017, Nature Astronomy, 1, 0117,
  \dodoi{10.1038/s41550-017-0117}

\bibitem[{{Jones} {et~al.}(2017){Jones}, {Van Winckel}, {Aller}, {Exter}, \&
  {De Marco}}]{jones2017}
{Jones}, D., {Van Winckel}, H., {Aller}, A., {Exter}, K., \& {De Marco}, O.
  2017, \aap, 600, L9, \dodoi{10.1051/0004-6361/201730700}

\bibitem[{{Jones} {et~al.}(2023){Jones}, {{\'A}lvarez-M{\'a}rquez}, {Sloan},
  {Kavanagh}, {Argyriou}, {Law}, {Labiano}, {Patapis}, {Mueller}, {Larson},
  {Bright}, {Klaassen}, {Fox}, {Gasman}, {Geers}, {Glauser}, {Guillard},
  {Nayak}, {Noriega-Crespo}, {Ressler}, {Sargent}, {Temim}, {Vandenbussche}, \&
  {Garc{\'\i}a Mar{\'\i}n}}]{jones2023}
{Jones}, O.~C., {{\'A}lvarez-M{\'a}rquez}, J., {Sloan}, G.~C., {et~al.} 2023,
  \mnras, 523, 2519, \dodoi{10.1093/mnras/stad1609}

\bibitem[{Law {et~al.}(2023)Law, Morrison, Argyriou, Patapis, Álvarez
  Márquez, Labiano, \& Vandenbussche}]{law2023}
Law, D.~R., Morrison, J.~E., Argyriou, I., {et~al.} 2023, The Astronomical
  Journal, 166, 45, \dodoi{10.3847/1538-3881/acdddc}

\bibitem[{{Luridiana} {et~al.}(2015){Luridiana}, {Morisset}, \& {Shaw}}]{pyneb}
{Luridiana}, V., {Morisset}, C., \& {Shaw}, R.~A. 2015, \aap, 573, A42,
  \dodoi{10.1051/0004-6361/201323152}

\bibitem[{{Masci} \& {Fowler}(2009)}]{masci2009}
{Masci}, F.~J., \& {Fowler}, J.~W. 2009, in Astronomical Society of the Pacific
  Conference Series, Vol. 411, Astronomical Data Analysis Software and Systems
  XVIII, ed. D.~A. {Bohlender}, D.~{Durand}, \& P.~{Dowler}, 67,
  \dodoi{10.48550/arXiv.0812.4310}

\bibitem[{{Mata} {et~al.}(2016){Mata}, {Ramos-Larios}, {Guerrero},
  {Nigoche-Netro}, {Toal{\'a}}, {Fang}, {Rubio}, {Kemp}, {Navarro}, \&
  {Corral}}]{mata2016}
{Mata}, H., {Ramos-Larios}, G., {Guerrero}, M.~A., {et~al.} 2016, \mnras, 459,
  841, \dodoi{10.1093/mnras/stw646}

\bibitem[{Meaburn {et~al.}(2005)Meaburn, Boumis, López, Harman, Bryce, Redman,
  \& Mavromatakis}]{meaburn2005}
Meaburn, J., Boumis, P., López, J.~A., {et~al.} 2005, Monthly Notices of the
  Royal Astronomical Society, 360, 963,
  \dodoi{10.1111/j.1365-2966.2005.09083.x}

\bibitem[{{Muthu} \& {Anandarao}(2003)}]{muthu2003}
{Muthu}, C., \& {Anandarao}, B.~G. 2003, \aj, 126, 2963, \dodoi{10.1086/379552}

\bibitem[{{Otsuka} {et~al.}(2017){Otsuka}, {Ueta}, {van Hoof}, {Sahai},
  {Aleman}, {Zijlstra}, {Chu}, {Villaver}, {Leal-Ferreira}, {Kastner},
  {Szczerba}, \& {Exter}}]{otsuka2017}
{Otsuka}, M., {Ueta}, T., {van Hoof}, P. A.~M., {et~al.} 2017, \apjs, 231, 22,
  \dodoi{10.3847/1538-4365/aa8175}

\bibitem[{{Parker} {et~al.}(2016){Parker}, {Boji{\v{c}}i{\'c}}, \&
  {Frew}}]{parker2016}
{Parker}, Q.~A., {Boji{\v{c}}i{\'c}}, I.~S., \& {Frew}, D.~J. 2016, in Journal
  of Physics Conference Series, Vol. 728, Journal of Physics Conference Series
  (IOP), 032008, \dodoi{10.1088/1742-6596/728/3/032008}

\bibitem[{{Ressler} {et~al.}(2010){Ressler}, {Cohen}, {Wachter}, {Hoard},
  {Mainzer}, \& {Wright}}]{ressler2010}
{Ressler}, M.~E., {Cohen}, M., {Wachter}, S., {et~al.} 2010, \aj, 140, 1882,
  \dodoi{10.1088/0004-6256/140/6/1882}

\bibitem[{{Stanghellini} {et~al.}(2012){Stanghellini},
  {Garc{\'\i}a-Hern{\'a}ndez}, {Garc{\'\i}a-Lario}, {Davies}, {Shaw},
  {Villaver}, {Manchado}, \& {Perea-Calder{\'o}n}}]{stang2012}
{Stanghellini}, L., {Garc{\'\i}a-Hern{\'a}ndez}, D.~A., {Garc{\'\i}a-Lario},
  P., {et~al.} 2012, \apj, 753, 172, \dodoi{10.1088/0004-637X/753/2/172}

\bibitem[{{Steffen} {et~al.}(2011){Steffen}, {Koning}, {Wenger}, {Morisset}, \&
  {Magnor}}]{steffen2011}
{Steffen}, W., {Koning}, N., {Wenger}, S., {Morisset}, C., \& {Magnor}, M.
  2011, IEEE Transactions on Visualization and Computer Graphics, 17, 454,
  \dodoi{10.1109/TVCG.2010.62}

\bibitem[{{Su} {et~al.}(2007){Su}, {Chu}, {Rieke}, {Huggins}, {Gruendl},
  {Napiwotzki}, {Rauch}, {Latter}, \& {Volk}}]{su2007}
{Su}, K.~Y.~L., {Chu}, Y.~H., {Rieke}, G.~H., {et~al.} 2007, \apjl, 657, L41,
  \dodoi{10.1086/513018}

\bibitem[{{Tinyanont} {et~al.}(2019){Tinyanont}, {Millar-Blanchaer}, {Nilsson},
  {Mawet}, {Knutson}, {Kataria}, {Vasisht}, {Henderson}, {Matthews}, {Serabyn},
  {Milburn}, {Hale}, {Smith}, {Vissapragada}, {Santos}, {Kekas}, \&
  {Escuti}}]{wircpol}
{Tinyanont}, S., {Millar-Blanchaer}, M.~A., {Nilsson}, R., {et~al.} 2019,
  \pasp, 131, 025001, \dodoi{10.1088/1538-3873/aaef0f}

\bibitem[{Virtanen {et~al.}(2020)Virtanen, Gommers, Oliphant, Haberland, Reddy,
  Cournapeau, Burovski, Peterson, Weckesser, Bright, {van der Walt}, Brett,
  Wilson, Millman, Mayorov, Nelson, Jones, Kern, Larson, Carey, Polat, Feng,
  Moore, {VanderPlas}, Laxalde, Perktold, Cimrman, Henriksen, Quintero, Harris,
  Archibald, Ribeiro, Pedregosa, {van Mulbregt}, \& {SciPy 1.0
  Contributors}}]{scipy2020}
Virtanen, P., Gommers, R., Oliphant, T.~E., {et~al.} 2020, Nature Methods, 17,
  261, \dodoi{10.1038/s41592-019-0686-2}

\bibitem[{{Wilson} {et~al.}(2003){Wilson}, {Eikenberry}, {Henderson},
  {Hayward}, {Carson}, {Pirger}, {Barry}, {Brandl}, {Houck}, {Fitzgerald}, \&
  {Stolberg}}]{wirc}
{Wilson}, J.~C., {Eikenberry}, S.~S., {Henderson}, C.~P., {et~al.} 2003, in
  Society of Photo-Optical Instrumentation Engineers (SPIE) Conference Series,
  Vol. 4841, Instrument Design and Performance for Optical/Infrared
  Ground-based Telescopes, ed. M.~{Iye} \& A.~F.~M. {Moorwood}, 451--458,
  \dodoi{10.1117/12.460336}

\bibitem[{{Zhang}(1995)}]{zhang1995}
{Zhang}, C.~Y. 1995, \apjs, 98, 659, \dodoi{10.1086/192173}

\bibitem[{{Zhang} \& {Kwok}(1993)}]{zhang1993}
{Zhang}, C.~Y., \& {Kwok}, S. 1993, \apjs, 88, 137, \dodoi{10.1086/191818}

\end{thebibliography}
\bibliographystyle{aasjournal}

\end{document}